\documentclass[lettersize,journal]{IEEEtran}
\usepackage{cite}
\usepackage{amsmath,amssymb,amsfonts}
\usepackage{algorithmic}
\usepackage{graphicx}
\usepackage{textcomp}
\usepackage{xcolor}
\usepackage{hyperref}
\usepackage{enumitem}
\usepackage{subfigure}
\usepackage{booktabs}

\def\BibTeX{{\rm B\kern-.05em{\sc i\kern-.025em b}\kern-.08em
    T\kern-.1667em\lower.7ex\hbox{E}\kern-.125emX}}
\begin{document}

\DeclareRobustCommand{\loongrightarrow}{%
  \DOTSB\relbar\joinrel\relbar\joinrel\rightarrow
}
\DeclareRobustCommand{\loongmapsto}{\DOTSB\mapstochar\loongrightarrow}

\DeclareRobustCommand{\looongrightarrow}{%
  \DOTSB\relbar\joinrel\relbar\joinrel\relbar\joinrel\rightarrow
}
\DeclareRobustCommand{\looongmapsto}{\DOTSB\mapstochar\looongrightarrow}

\title{AI-Empowered Low-Altitude Economy: Cooperative Sensing With Fixed Wireless Access}
\author{Jinya~Zhang,~Jiajia~Guo,~Xiangyi~Li,~Chao-Kai~Wen, \IEEEmembership{\normalsize{Fellow,~IEEE}}, and Shi~Jin, \IEEEmembership{\normalsize{Fellow,~IEEE}}

\thanks{This work has been partly accepted by IEEE ICC Workshops 2026\cite{zhang2026icc}.}% <-this % stops a space
}

\maketitle

\begin{abstract}
The rapid growth of the low-altitude economy has intensified safety concerns arising from unauthorized unmanned aerial vehicles (UAVs), positioning UAV supervision as a key use case in 3GPP. To precisely sense such UAVs with wide coverage and low cost, we leverage fixed wireless access (FWA) customer premises equipment (CPEs), static, densely deployed devices that serve as wireless cameras for the radio environment. 
We develop an artificial intelligence-empowered two-stage cooperative sensing pipeline that exploits uplink channel state information (CSI) from multiple base station-CPE pairs for UAV detection and localization. In cooperative detection, lightweight CSI features are first individually extracted by neural network, and then adaptively integrated through an attention-based scheme to declare UAV presence. The learned attention scores effectively identify the critical pairs during detection, while facilitating UAV-affected pair selection for subsequent localization. For cooperative localization, neural network initially generates individual estimates and extract CSI features from selected pairs.
These estimates, together with features and pair indexes, are fused using a Transformer to produce a precise cooperative estimate. 
Simulations show that cooperative schemes significantly reduce the missed detection probability to 0.63\% and realize a 95\%-confidence positioning error of 6.50~m, satisfying 3GPP requirements and showing the potential of FWA-assisted cooperative sensing. Dataset and codes are available on GitHub\footnote{The GitHub link will be provided once the paper is accepted. https://github.com/xxxxxx.}.
\end{abstract}


\section{Introduction}
\IEEEPARstart{T}HE low-altitude economy (LAE), typically defined as economy activities in airspace below 3,000~m, has seen rapid deployment of unmanned aerial vehicles (UAVs) for commercial, inspection, and logistics applications \cite{yuan2025ground, xu2025llmlae}. Nevertheless, unauthorized UAVs, or black flights, pose serious threats to infrastructure, public safety, and privacy, creating an urgent demand for reliable detection and localization \cite{wu2023precise}. Addressing this challenge requires new paradigms beyond conventional monitoring systems. As a core usage scenario in the 6th generation mobile communication systems (6G) vision \cite{ITU2022}, integrated sensing and communication (ISAC) provides a promising solution for co-locating sensing and communication functions within the same infrastructure \cite{liu2022integrated,elfiatoure2025multi,fu2025beam}. By reusing existing wireless networks, radio-signal-based sensing enables cost efficiency, ubiquitous coverage, and non-intrusiveness \cite{zhang2022practical}. Reflecting its relevance, 3GPP TR 22.870 \cite{3GPP22870} (Release 20 study on 6G use cases) has explicitly recognized low-altitude UAV supervision as a representative ISAC use case. 

\begin{figure}[!t]
    \centering
    \includegraphics[width=1\linewidth]{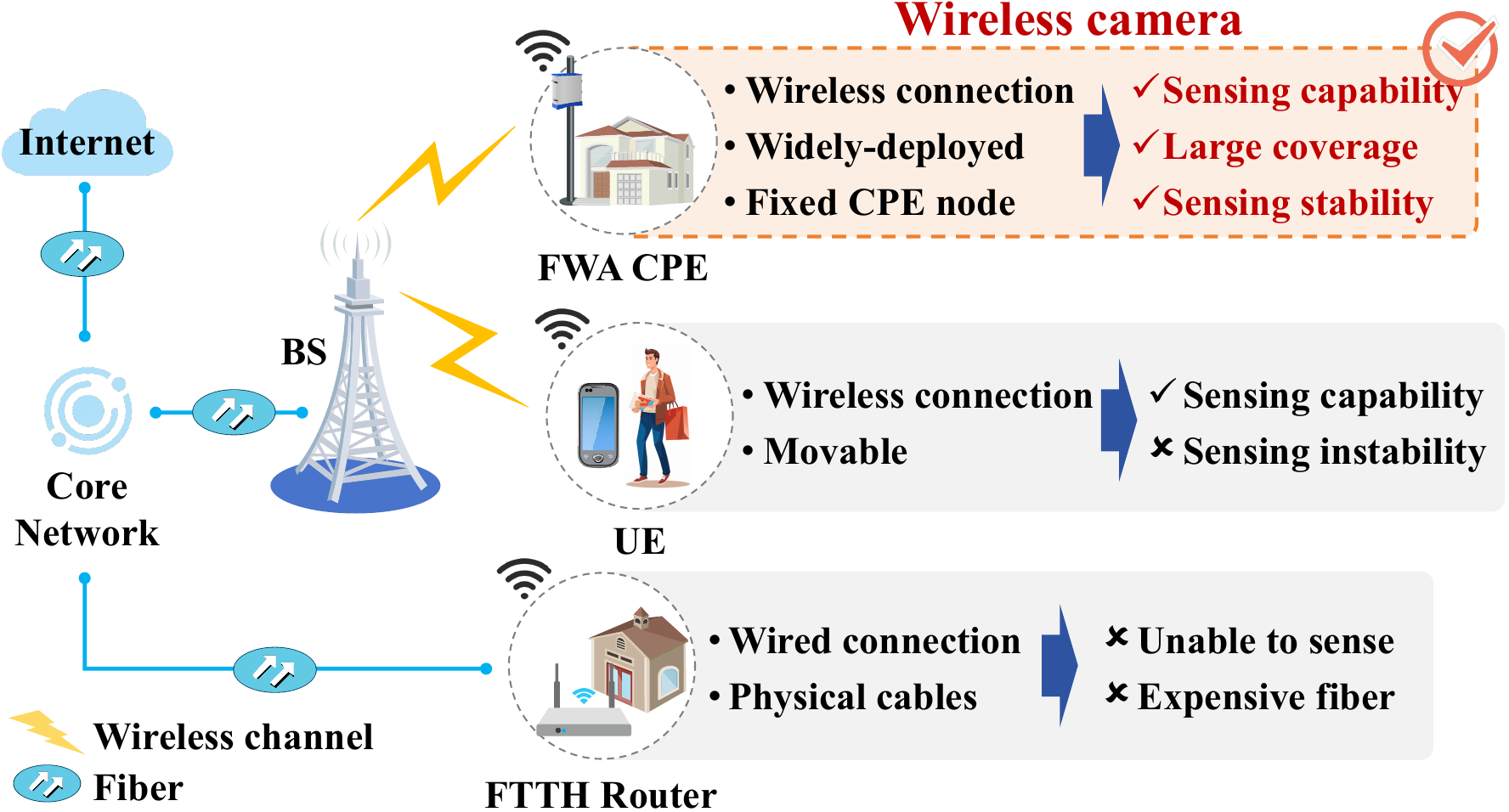}
    \caption{Widely-deployed, fixed FWA CPEs show wireless sensing potential.}
    \label{fig:FWA}
    \vspace{-0.4cm}
\end{figure}

Since black flights do not disclose onboard information (e.g., GNSS coordinates), they must be sensed under device-free paradigms in which targets do not cooperate with the supervisor \cite{wang2018device}. In ISAC-enabled device-free sensing, commonly used sensing metrics include received signal strength (RSS), time of arrival (ToA), angle of arrival (AoA), and channel state information (CSI) \cite{zhang2022toward}. As a measure of signal amplitude, RSS can indicate object presence or provide coarse range estimates via path loss models, but its reliability deteriorates in multipath-rich environments \cite{Konings2019rssi}. ToA infers distance from propagation delay, but requires substantial bandwidth to attain fine temporal resolution \cite{shen2012toa}. Likewise, AoA supplies directional information that facilitates localization, yet achieving high angular resolution necessitates large antenna arrays \cite{zhang2018aoa}. In contrast to metrics capturing only partial information, CSI encodes finer-grained propagation characteristics (e.g., multipath signatures), which are more informative for environment inference \cite{guo2024learning, ma2019wifi, zhang2023multi,meneghello2022sharp}. Additionally, artificial intelligence (AI)-based methods have shown the ability to exploit rich features \cite{wei2022csi, guo2026prompt}, yielding stronger robustness and higher accuracy compared to traditional model-based approaches \cite{shi2022device,li2016dynamic}. However, many CSI-based works rely on a solitary transmitter-receiver pair, causing limited coverage and vulnerability in multipath-rich or obstructed channels. 

To surmount these limitations, node cooperation (fusion of multiple sensing observations or decisions) has been adopted to enhance sensing reliability \cite{wang2025cooperative}. Early efforts have mainly focused on cooperative frameworks involving multiple base stations (BSs). \cite{tang2025cooperative} has integrated multiple monostatic BSs into a two-stage cooperative UAV sensing pipeline. Within a similar scenario, \cite{figueroa2024optimal} has incorporated ToA and AoA measurements from multiple monostatic BSs with optimally chosen weights to refine positioning accuracy. Subsequent studies have extended BS cooperation architectures with diverse sensing equipment. For instance, the authors of \cite{wang2025isacen} have leveraged a cellular-connected UAV to assist the sensing process, and reconfigurable intelligent surfaces (RIS) have been employed as passive anchors in \cite{wang2024heterogeneous}. Beyond framework design, transmission overhead has remained a critical concern. To address this, a codebook-based sensing data transmission scheme has been exploited for efficient cooperation between multiple access points \cite{wang2025cojsac}. Collectively, these cooperative sensing works substantially mitigate the coverage and robustness limitations compared to single-pair sensing.

For cooperative sensing, sensing node selection, effective data processing and fusion, and efficient data transmission are three essential components. Among these, node selection serves as the foundational element. In LAE supervision, where the monitored airspace spans large regions, strategic node selection is particularly critical to ensure seamless and reliable sensing. Although BSs achieve high sensing accuracy owing to high-end radios and wide bandwidth, their sparse deployment constrains spatial coverage \cite{liu2024leveraging}. Augmenting BS-centric systems with widely deployed user equipment (UEs) can increase coverage, but UE mobility undermines sensing stability. RISs help establish line-of-sight (LoS) links, but their limited signal processing capabilities complicate the sensing process. Under these considerations, a critical challenge in reliable, large-area LAE supervision is:

\vspace{1mm}
\noindent\emph{Are there alternative widely-deployed fixed nodes to assist BS-centric sensing?}
\vspace{1mm}

The rapid development of fixed wireless access (FWA) \cite{ericssonFWA} has catalyzed a dense deployment of customer premises equipment (CPE). As illustrated in Fig.~\ref{fig:FWA}, these devices possess several advantages for sensing, including fixed placement, high spatial density, and persistent connectivity with nearby BSs. This observation motivates a reconsideration of the role of CPEs: beyond functioning as last-mile communication devices, they can potentially be repurposed as wireless cameras for the radio environment, delivering rich, fine-grained observations suitable for cooperative sensing. While the 3GPP has recognized the potential of CPEs for indoor sensing tasks \cite{3GPP22137,3GPP22837}, extending their application to LAE supervision presents several challenges, including ensuring seamless coverage over wide areas, accommodating heterogeneous CSI statistics, and minimizing sensing data transmission overhead. According to these considerations, we leverage FWA to assist BS-centric sensing for LAE supervision and investigate an effective sensing framework to tackle these challenges.

Motivated by the demand for accurate, wide-area, and cost-effective monitoring in LAE, and enabled by the potential of FWA-assisted sensing in ISAC, we propose an AI-based two-stage cooperative sensing pipeline that leverages uplink CSI from multiple BS-CPE pairs to detect and localize the unauthorized UAV. The key contributions of this study are summarized as follows: 
\begin{itemize}[leftmargin=*]
    \item \textbf{FWA-assisted sensing framework.} We introduce the novel use of static, densely deployed FWA CPEs as auxiliary sensing nodes, exploiting their fixed placement and high spatial density to complement BS-centric sensing by functioning as wireless cameras. Sensing region visualizations confirm that integrating multiple BSs and CPEs substantially eliminates blind spots and expands effective coverage.

    \item \textbf{AI-based cooperative detection architecture.} We design a two-stage framework following the attention-based multiple instance learning (MIL) paradigm. First, lightweight CSI features are individually extracted using neural network (NN) at BSs and transmitted to central processing unit (CPU). Second, these features are cooperatively integrated to infer UAV presence, with attention scores assigned to reflect the relative importance of different pairs in the detection process. In general, higher attention scores are typically assigned to UAV-affected pairs. By transmitting only CSI features, this two-stage design significantly reduces BS-CPU transmission overhead.
    
    \item \textbf{AI-based cooperative localization architecture.} We further develop a two-stage localization framework. Initially, UAV-affected pairs are selected under the guidance of attention scores learned from detection, and only the selected pairs participate in localization. In the first stage, individual localization NN predicts UAV positions and extract CSI features. These coarse estimates, along with associated features and pair indexes, are subsequently transmitted and fused at the CPU to produce a refined estimate using a cooperative NN. This lightweight fusion strategy effectively balances transmission overhead and localization accuracy.
    \item \textbf{Performance evaluation and insights:} Simulations under ray-tracing channel models demonstrate that the FWA-aided cooperative detection scheme significantly extends sensing coverage and reduces the missed detection probability (MDP) to 0.63\%, while the cooperative localization scheme achieves an accuracy of positioning error (APE) of 6.50~m at a 95\% confidence level, both satisfying 3GPP requirements. In-depth analyses further explore trade-offs in attention thresholds, feature extraction, node deployment density, and robustness under cluttered propagation conditions.
\end{itemize}

The rest of this paper is organized as follows. Section II introduces the LAE scenario and system model. Section III details the AI-based cooperative UAV detection framework. Section IV presents the AI-based cooperative UAV localization design. Section V provides simulation settings, results, and analysis. Finally, Section VI concludes the paper.

\textit{Notations:} Lowercase symbols denote scalars, bold lowercase symbols denote column vectors, and bold uppercase symbols denote matrices. The $\odot$ and $\land$ represent the element-wise multiplication and the logical AND operation, respectively. $X\sim \mathcal{U}(a,b)$ denotes a uniform distribution between $a$ and $b$. The notation $(\cdot)^\top$ is the transposition of a matrix.

\section{System Model}
\begin{figure}[!t]
    \centering
    \includegraphics[width=1\linewidth]{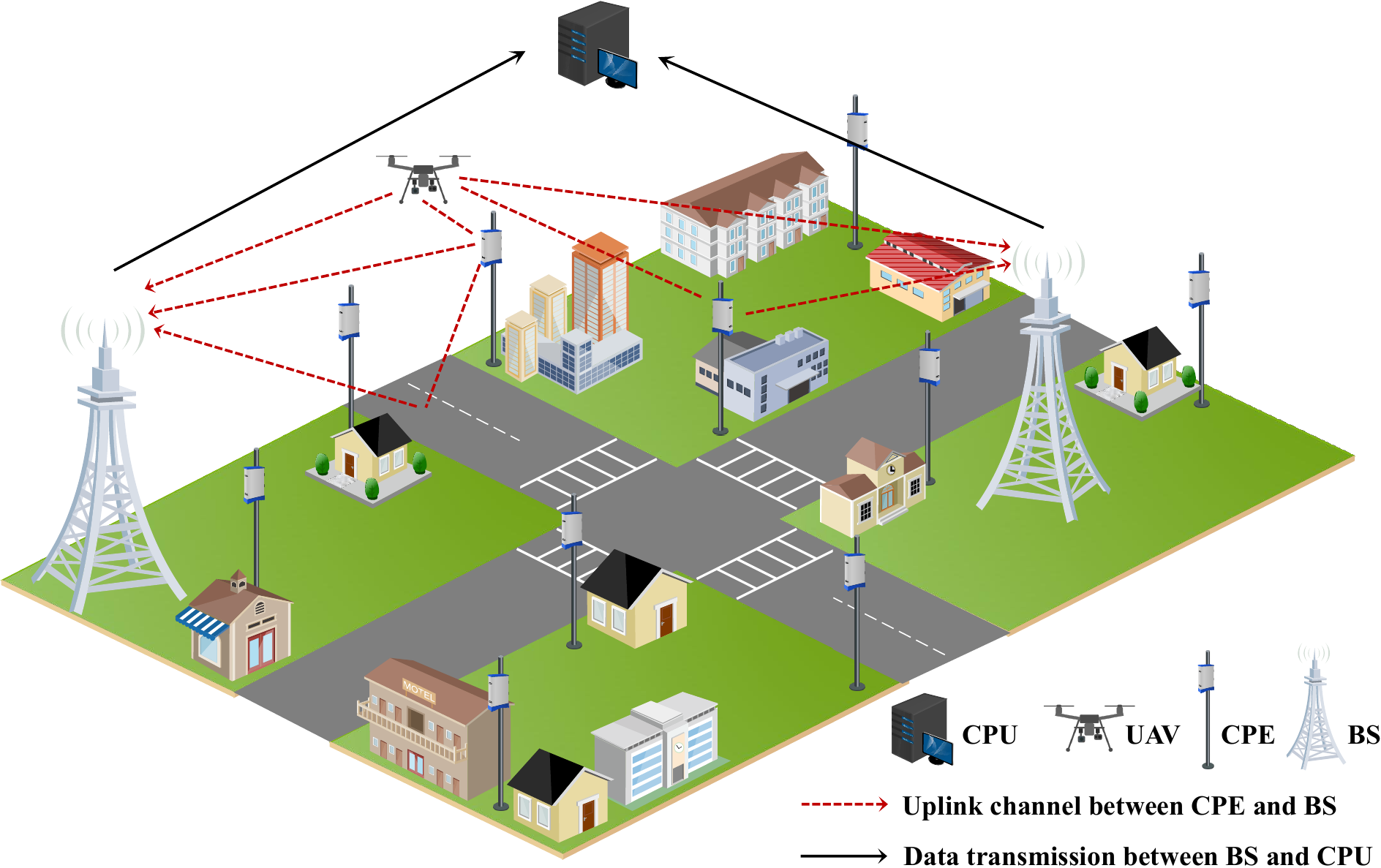}
    \caption{The LAE scenario comprises a CPU, multiple BSs/CPEs, and a UAV.}
    \label{fig:LAEscenario}
\end{figure}

This work considers uplink transmission in a multiple-input multiple-output orthogonal frequency division multiplexing (MIMO-OFDM) system deployed in a three-dimensional LAE scene. The scenario consists of a CPU, $M$ BSs, $N$ CPEs, and a single UAV, as illustrated in Fig.~\ref{fig:LAEscenario}. BSs and CPEs are static and are located at $(x^m_\mathrm{bs}, y^m_\mathrm{bs}, z^m_\mathrm{bs})$, $m=1,\dots, M$, and $(x^n_\mathrm{cpe}, y^n_\mathrm{cpe}, z^n_\mathrm{cpe})$, $n=1, \dots, N$, respectively. The UAV hovers at $\mathbf{p}=(x,y,z)$ and acts as a scatterer whose presence affects the wireless propagation.

Each BS is equipped with $N_\mathrm{r}$ antennas and CPE with $N_\mathrm{t}$ antennas. Both of them adopt uniform plane arrays (UPA) with half-wavelength spacing. The system employs $N_\mathrm{c}$ OFDM subcarriers. All $N$ CPEs can communicate with all $M$ BSs, yielding $M \times N$ BS-CPE pairs in total. Through pilot transmission, BSs acquire uplink CSI on the space-frequency grid for every BS-CPE pair. For simplicity, we assume synchronous CSI acquisition across all BSs on the same set of subcarriers and neglect mutual interference. Accordingly, the collected channel frequency responses (CFRs) are represented as $\mathbf{H}^{(1,1)},\dots,\mathbf{H}^{(M, N)}$, where
\begin{equation}
    \mathbf{H}^{(m,n)} = \left[\mathbf{H}_1^{(m,n)},\dots,\mathbf{H}_{N_\mathrm{c}}^{(m,n)} \right] \in \mathbb{C}^{N_\mathrm{r} \times N_\mathrm{t} \times N_\mathrm{c}}.
\end{equation}

To relate the CFR to the physical environment, the wireless channel is modeled as a superposition of multipath components. Let $i=1, \dots, N_\mathrm{r}$ and $j=1, \dots, N_\mathrm{t}$ index receiving and transmitting antennas, respectively. The channel at frequency $f$ can be written as
\begin{equation}
    H_{i,j}(f) = \sum_{p=1}^{N_{\mathrm{p}}} \alpha_{i,j,p} e^{-j2\pi f \tau_{p}},
    \label{eq:ori_channel}
\end{equation}
where $N_\mathrm{p}$ is the number of propagation paths, $\alpha_{i,j,p}$ is the complex gain of path $p$, and $\tau_{p}$ is its delay. The presence of a UAV inside the sensing region modifies multipath propagation for several BS-CPE pairs. Consequently, the $M \times N$ collection of CSI measurements can be exploited to infer the presence and position of a UAV using the existing infrastructure.

\section{AI-Based Cooperative UAV Detection}
\begin{figure*}[!t]
    \centering
    \includegraphics[width=1\linewidth]{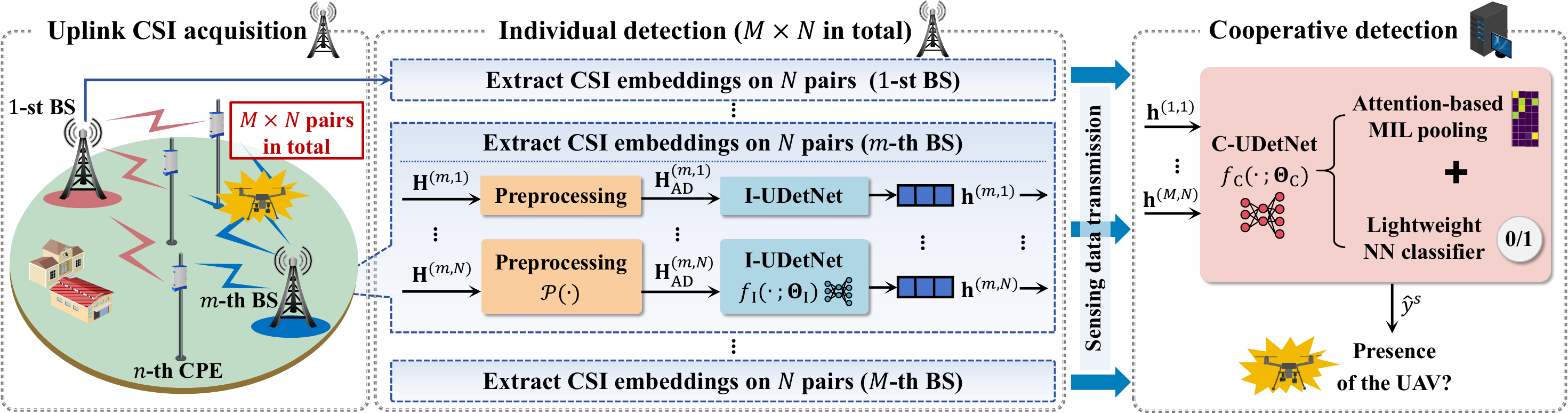}
    \caption{The two-stage cooperative UAV detection framework. In \textbf{Stage 1 (individual detection)}, each BS preprocesses $N$ uplink CSI and generates compact embeddings that preserve UAV-informative features. In \textbf{Stage 2 (cooperative detection)}, CSI embeddings are transmitted to the CPU and fused with an attention-based mechanism to yield the final detection.}
    \label{fig:UAVdetection}
\end{figure*}

In this section, we propose an AI-based two-stage cooperative UAV detection framework using the CSI of multiple BS-CPE pairs. The framework follows the MIL paradigm \cite{ilse2018attention}, as shown in Fig.~\ref{fig:UAVdetection}. Specifically, each BS acquires uplink CSI from $N$ CPEs, collecting $M \times N$ CSI measurements. First, AI-based individual detection is performed on each BS-CPE pair to encode lightweight CSI embedding, which are designed to extract information associated with UAV-affected channel variations. Second, all CSI embeddings are fused through an attention-based mechanism to produce a cooperative detection result that identifies the presence of UAV.

\subsection{Overall UAV Detection Framework Based on MIL}
In UAV detection tasks, CSI measurements collected from multiple BS-CPE pairs provide a set of multi-view observations of the monitored LAE airspace. Since the impact of a UAV on wireless propagation depends on the surrounding geometry, and each BS-CPE pair has only a limited sensing region, different pairs may respond quite differently to the UAV. As a result, only a subset of the BS-CPE pairs exhibit evident UAV-related channel variations, while others may remain weakly affected or unaffected. Moreover, although the scene label, i.e., whether a UAV is present in the airspace, can be obtained, identifying which specific BS-CPE pairs are affected by the UAV is hard in practical deployments.

Such a setting naturally matches the MIL paradigm. MIL is a weakly supervised learning framework in which the training data are organized as bags, each consisting of multiple instances, and only bag labels are available. Instead of requiring explicit supervision for every instance, MIL learns to infer the bag label by identifying the informative instances relevant to the target task. Accordingly, in the UAV detection task, each monitored scene is treated as a bag, and the CSI measurements collected from all BS-CPE pairs are regarded as the instances within that bag. The corresponding bag label (or scene label), denoted by $y^\mathrm{s}\in\{0,1\}$, indicates whether a UAV is present in the LAE airspace. Assume that the instance labels (or pair labels) $Y=\{y^{(1,1)},\dots,y^{(M,N)}\}$ exist but are inaccessible, where $y^{(m,n)}\in\{0,1\}$ indicates whether the CSI of the $(m,n)$-th BS-CPE pair is affected by the UAV. Consequently, the scene label can be determined from all pair labels as
\begin{equation}
y^\mathrm{s}
=
\begin{cases}
0, & \sum^{M}_{m=1}\sum^{N}_{n=1}{y^{(m,n)}} = 0,\\[4pt]
1, & \text{otherwise}.
\end{cases}
\end{equation}

The unavailability of pair labels limits the applicability of the pair-supervised two-stage detection strategy proposed in \cite{zhang2026icc}. By contrast, MIL provides a more suitable formulation, as it predicts the presence of UAV using only scene supervision. In particular, the attention-based MIL model in \cite{ilse2018attention} jointly learns instance representations and bag-level aggregation, while adaptively emphasizing the BS-CPE pairs containing discriminative UAV-related information. Therefore, it yields reliable and interpretable cooperative detection results. The detailed framework is presented as follows.

\subsection{Stage 1: Individual Detection}
\begin{figure*}[!t]
    \centering
    \includegraphics[width=0.95\linewidth]{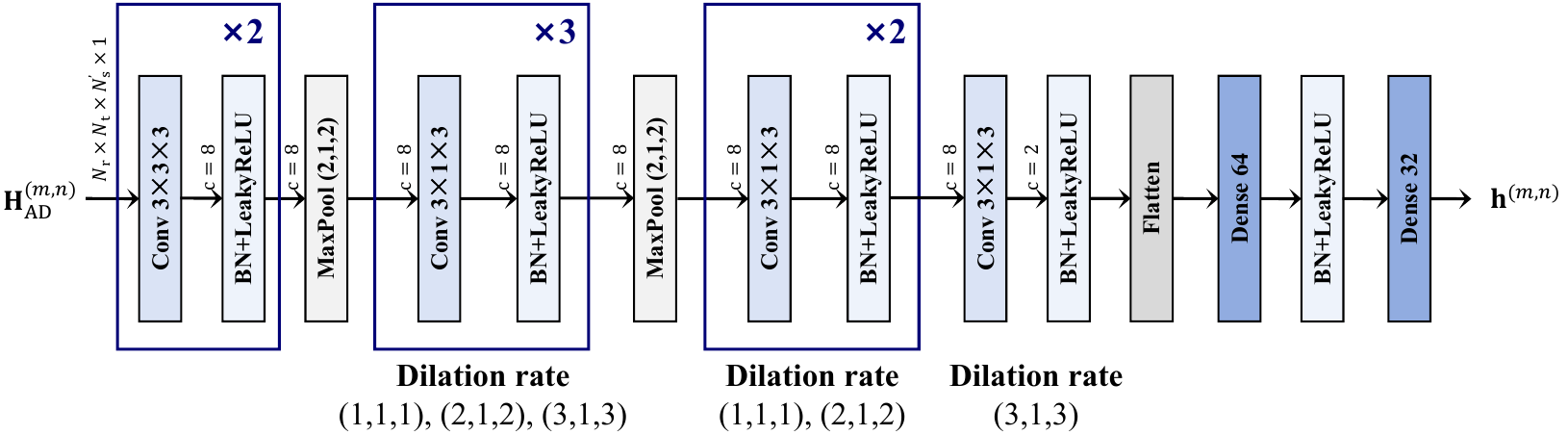}
    \caption{The NN architecture of the I-UDetNet for individual detection.}
    \label{fig:I-UDetNet}
\end{figure*}
The individual detection stage aims to extract UAV-related information and produce lightweight CSI embeddings $\mathbf{h}^{(m,n)} \in \mathbb{R}^{N_\mathrm{e}}$ based on each CSI $\mathbf{H}^{(m,n)}$, where $N_\mathrm{e}$ is the embedding length. Individual detection comprises two modules: CSI preprocessing and a neural encoder, denoted as \underline{I}ndividual \underline{U}AV \underline{Det}ection \underline{Net} (I-UDetNet). For the $(m,n)$-th BS-CPE pair, individual detection is formulated as 

\begin{equation}
    \mathbf{h}^{(m,n)} = f_\mathrm{I}(\mathcal{P}(\mathbf{H}^{(m,n)});\mathbf{\Theta}_\mathrm{I}),
\end{equation}
where $\mathcal{P}(\cdot)$ represents the preprocessing operator, $f_\mathrm{I}(\cdot)$ and $\mathbf{\Theta}_\mathrm{I}$ denotes the I-UDetNet and its parameters, respectively.

\subsubsection{CSI Preprocessing}
\label{subsubsec:CSIpre}
Compared to the space-frequency domain, expressing CFR in the angle-delay domain provides a more physically interpretable representation that is appropriate to UAV sensing. Therefore, we convert the CFR $\mathbf{H}^{(m,n)}$ to the angle-delay channel response

\begin{equation}
    \mathbf{H}_\mathrm{AD}^{(m,n)} = \mathcal{P}(\mathbf{H}^{(m,n)}),
\end{equation}
where $\mathbf{H}_\mathrm{AD}^{(m,n)} \in \mathbb{R}^{N_\mathrm{r} \times N_\mathrm{t} \times N_{\mathrm{c}}' \times 1}$. The operator $\mathcal{P}(\cdot)$ includes four steps, which are detailed below.

\begin{enumerate}[leftmargin=*,label=\textbf{\alph*.}]
    \item \textbf{3D-IDFT.} The first step is to apply a three-dimensional inverse discrete Fourier transform (3D-IDFT) along the antenna and subcarrier axes
    \begin{equation}
        \mathbf{H}^{(m,n)} \stackrel{\text{3D-IDFT}}{\looongrightarrow} \widetilde{\mathbf{H}}_{\mathrm{AD}}^{(m,n)},
    \end{equation}
    where $\widetilde{\mathbf{H}}_{\mathrm{AD}}^{(m,n)} \in \mathbb{C}^{N_\mathrm{r} \times N_\mathrm{t} \times N_\mathrm{c}}$.

    \item \textbf{Truncation.} Since the delay energy is typically gathered in a limited period, we retain the first $N_{\mathrm{c}}'$ delay elements and discard the remainder to reduce input dimensionality
    \begin{equation}
    \widetilde{\mathbf{H}}_{\mathrm{AD}}^{(m,n)}(:,: , 1\!:\!N_{\mathrm{c}}') \rightarrow \widetilde{\mathbf{H}}_{\mathrm{AD}}^{(m,n)} \in\mathbb{C}^{N_{\mathrm{r}}\times N_{\mathrm{t}}\times N_{\mathrm{c}}'}.
    \end{equation}

    \item \textbf{Amplitude extraction and logarithmic compression.} We extract the magnitude of $\widetilde{\mathbf{H}}_{\mathrm{AD}}^{(m,n)}$ and apply a logarithmic transform to compress the dynamic range, which helps reveal low-energy paths affected by the UAV
    \begin{equation}
        \mathbf{H}_{\mathrm{ADAMP}}^{(m,n)} = \log_{10}(|\widetilde{\mathbf{H}}_{\mathrm{AD}}^{(m,n)}|), 
        \label{eq:prepro_log}
    \end{equation}
    where $\mathbf{H}_{\mathrm{ADAMP}}^{(m,n)} \in \mathbb{R}^{N_\mathrm{r} \times N_\mathrm{t} \times N_{\mathrm{c}}'}$.
    
    \item \textbf{Reshape and min-max normalization.} Finally, $\mathbf{H}_{\mathrm{ADAMP}}^{(m,n)}$ is 
    reshaped to add a channel dimension to match the input shape expected by convolutional (Conv) layers, and min-max normalization is applied (per sample) to scale the values into the interval $(0,1)$
    \begin{equation}
        \mathbf{H}_\mathrm{AD}^{(m,n)}= {\rm Norm}{\left( {\rm Reshape}(\mathbf{H}_{\mathrm{ADAMP}}^{(m,n)} )\right)}.
    \end{equation}
\end{enumerate}

\subsubsection{I-UDetNet Design}
Following CSI preprocessing, the I-UDetNet encodes each $\mathbf{H}_\mathrm{AD}^{(m,n)}$ into a compact embedding that preserves the CSI features most informative of UAV-induced perturbations. In the individual detection stage, a shared I-UDetNet for all BS-CPE pairs is sufficient.
 
\begin{itemize}[leftmargin=*]
    \item \textbf{Input and output.} The I-UDetNet takes the preprocessed CSI $\mathbf{H}^{(m,n)}_\mathrm{AD}$ for a single pair as input and produces a compact embedding  $\mathbf{h}^{(m,n)}$. This embedding is expected to preserve the CSI features that are most informative of UAV-induced channel variations. Since UAV may influence propagation paths compared with scenarios without UAV, $\mathbf{h}^{(m,n)}$ provides valuable evidence for the subsequent cooperative detection stage to infer the UAV presence.

    \item \textbf{NN architecture.} The NN architecture is shown in Fig.~\ref{fig:I-UDetNet}, comprising Conv3D, Batch Normalization (BN), MaxPool, Flatten, and Dense layers. Leaky ReLU and Sigmoid activations are used. To enlarge the receptive field without reducing spatial resolution, we employ multiple dilated Convs with different dilation rates \cite{yu2015multi}. UAV-affected scattering appears as sparse and spatially correlated energy across neighboring angle-delay bins, while dilated Convs help capture multi-scale patterns indicative of UAV presence.
    
\end{itemize}

\subsection{Stage 2: Cooperative Detection}
After obtaining $M \times N$ CSI embeddings, $\mathbf{h}^{(1,1)},\dots,\mathbf{h}^{(M,N)}$ are transmitted from BSs to the CPU, where the presence of a UAV is inferred through a cooperative detection process. This process is implemented by \underline{C}ooperative \underline{U}AV \underline{Det}ection \underline{Net} (C-UDetNet), consisting attention-based MIL pooling module \cite{ilse2018attention} and a lightweight classifier. The cooperative detection stage can be written as
\begin{equation}
    \hat{y}^\mathrm{s}=f_\mathrm{C}(\mathbf{h}^{(1,1)},\dots,\mathbf{h}^{(M,N)}; \mathbf{\Theta}_\mathrm{C}),
\end{equation}
where $f_\mathrm{C}(\cdot)$ and $\mathbf{\Theta}_\mathrm{C}$ denotes the C-UDetNet and its parameters, respectively, $\hat{y}^\mathrm{s} \in \{0,1\}$ represents the cooperative decision on the presence of UAV.

\begin{itemize}[leftmargin=*]
    \item \textbf{Attention-based MIL pooling.} Firstly, all CSI embeddings are aggregated into a single vector $\mathbf{z} \in \mathbb{R}^{N_\mathrm{e}}$. In this process, $M \times N$ attention weights are learned to adaptively measure the relative contribution of different BS-CPE pairs to UAV detection. The aggregated feature is given by
    \begin{equation}\mathbf{z}=\sum_{m=1}^M\sum_{n=1}^Na_{m,n}\mathbf{h}^{(m,n)},
    \end{equation}
    where the attention score of each pair is calculated as
    \begin{equation}
    \begin{gathered}
    s_{m,n} = \mathbf{w_\mathrm{a}}^\top \left(    \operatorname{Tanh}\!\left(\mathbf{V_\mathrm{a}}\mathbf{h}^{(m,n)}\right)
    \odot
    \operatorname{Sigm}\!\left(\mathbf{U_\mathrm{a}}\mathbf{h}^{(m,n)}\right)    \right), \\
    a_{m,n} = \frac{\exp(s_{m,n})} {\sum_{m=1}^{M}\sum_{n=1}^{N}\exp(s_{m,n})}.
    \label{eq:att_score}
    \end{gathered}
    \end{equation}
    Here, $\mathbf{w_\mathrm{a}}$, $\mathbf{V_\mathrm{a}}$, and $\mathbf{U_\mathrm{a}}$ denote the learnable parameters of three Dense layers. Through attention-based MIL pooling mechanism, C-UDetNet can assign larger weights to the pairs that strongly affected by UAV, while reducing the influence of weakly affected or unaffected BS-CPE pairs. As a result, the feature $\mathbf{z}$ provides a compact while valuable representation for cooperative UAV detection.
    \item \textbf{Lightweight classifier.} Based on the merged feature $\mathbf{z}$, a lightweight Dense classifier is employed to predict the UAV presence $\hat{y}^\mathrm{s}$. Specifically, the classifier consists of two Dense layers with ReLU and Sigmoid activations, respectively, together with a Dropout layer for regularization. Combined with the attention mechanism, the classifier enables the cooperative detection stage to remain computationally efficient while effectively exploiting the complementary sensing information provided by multiple BS-CPE pairs.
\end{itemize}

\subsection{Training Strategy and Evaluation Metrics}
\label{subsec:Training Strategy}
I-UDetNet and C-UDetNet are jointly optimized using ground-truth label $y^\mathrm{s}$ with binary cross-entropy loss, while no pair labels are required during training. Under this end-to-end training strategy, I-UDetNet is encouraged to extract CSI features that are beneficial for the subsequent cooperative decision, thereby improving the detection accuracy. Once offline training is completed, I-UDetNet is deployed at each BS to process the local CSI measurements of all associated BS-CPE pairs, and C-UDetNet is deployed at the CPU for cooperative detection. Each BS only needs to upload the lightweight CSI embeddings extracted by I-UDetNet, rather than high-dimensional CSI. This significantly reduces the transmission overhead between BSs and the CPU. Moreover, since C-UDetNet performs attention-based MIL pooling, the CPU can additionally obtain the attention score of each BS-CPE pair, providing useful auxiliary information for subsequent tasks, such as UAV localization. 

To improve the robustness of the detection framework against variations in the number of available BS-CPE pairs, a random pair-sampling strategy is introduced during training. Specifically, the CSI of all $M\times N$ pairs is not always provided for each training sample. Instead, only a randomly selected subset of pairs is retained, with the subset size uniformly sampled from $[\frac{M\times N}{2},\, M\times N]$. In this way, the network is prevented from over-relying on a few specific pairs. As a result, the proposed framework becomes more robust to pair missing or pair selection changes during inference.

For the two-stage detection pipeline, MDP and false alarm probability (FAP) \cite{3GPP22137} are adopted to indicate the ability to identify UAV's presence in the LAE scenario. MDP measures the likelihood of failing to detect a UAV when it is actually present, and is defined as
\begin{equation}
    \text{MDP} = \frac{\mathtt{FN}}{\mathtt{FN}+\mathtt{TP}}
    =\frac{(\hat{y}^\mathrm{s}=0) \land (y^\mathrm{s}=1)}{y^\mathrm{s}=1},
\end{equation}
where $\mathtt{FN}$ and $\mathtt{TP}$ denote the numbers of missed and correctly detected UAV-present samples, respectively. Consequently, $\mathtt{FN}+\mathtt{TP}$ represents the total number of samples in which a UAV is present.

Similarly, FAP quantifies the probability of falsely declaring a UAV when none exists, and is given by
\begin{equation}
    \text{FAP} = \frac{\mathtt{FP}}{\mathtt{FP}+\mathtt{TN}} =\frac{(\hat{y}^\mathrm{s}=1) \land (y^\mathrm{s}=0)}{y^\mathrm{s}=0},
\end{equation}
where $\mathtt{FP}$ and $\mathtt{TN}$ are the numbers of false positive and true negative samples, respectively. In this case, $\mathtt{FP}+\mathtt{TN}$ corresponds to the total number of UAV-absent samples.

\section{AI-Based Cooperative UAV Localization}
\begin{figure*}[!t]
    \centering
    \includegraphics[width=1\linewidth]{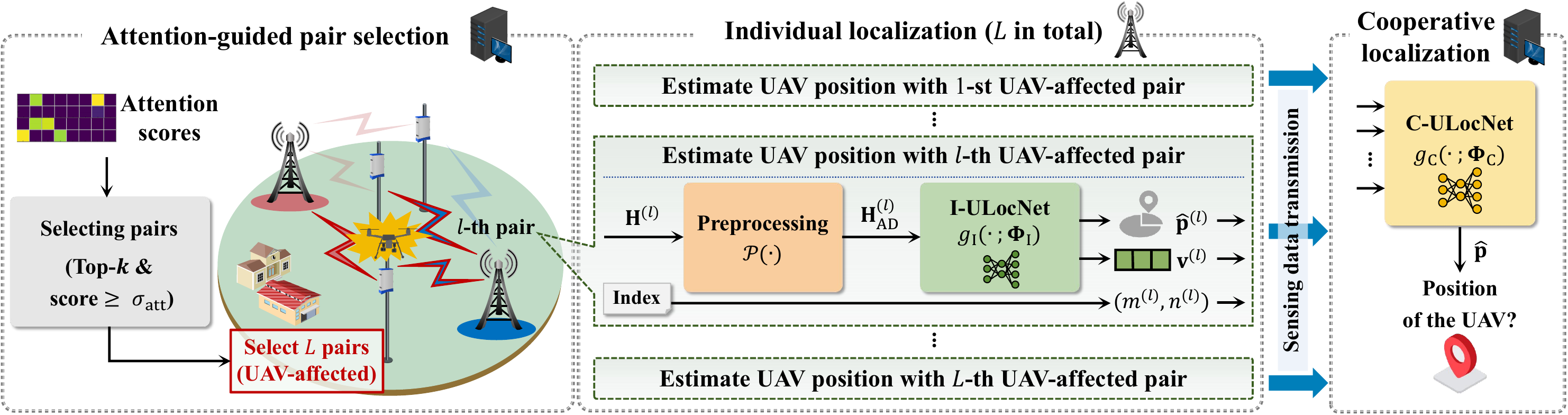}
    \caption{The two-stage cooperative UAV localization framework. Prior to localization, the CPU screens $L$ UAV-affected pairs according to the attention scores obtained in cooperative detection. In \textbf{Stage 1 (individual localization)}, the corresponding BSs are instructed to generate individual positioning estimates and extract CSI features from selected pairs. In \textbf{Stage 2 (cooperative localization)}, individual estimates, features, and pair indexes are fused at the CPU to produce a refined estimate.}
    \label{fig:UAVlocalization}
    \vspace{-0.3cm}
\end{figure*}
In this section, we further design an AI-based two-stage cooperative UAV localization framework. Specifically, a medium fusion strategy is adopted, which operates on intermediate representations, i.e., local estimates and CSI features. Compared with hard fusion, which uses only local estimates, and soft fusion, which directly fuses raw CSI, the adopted strategy provides a better tradeoff between sensing information retention and transmission cost. As illustrated in Fig.~\ref{fig:UAVlocalization}, the framework begins with identifying $L$ BS-CPE pairs according to the detection attention scores. The CSI of these pairs are expected to be more strongly affected by the UAV and thus suitable for position estimation. In the first stage, a localization NN estimates a coarse position and simultaneously extracts a compact CSI feature of each selected pair. These $L$ estimates, along with the corresponding features and pair indexes, are transmitted to the CPU for fusion. In the cooperative localization stage, the CPU employs a Transformer-based medium-fusion strategy to integrate these heterogeneous inputs and produce a refined positioning estimate. The detailed components of the framework are elaborated below.

\subsection{Preparation: Attention-Guided Pair Selection}
\label{sec:pairscreen}
Due to the limited sensing coverage of each BS-CPE pair, different pairs generally contribute unequally to UAV localization. Pairs whose channels are more strongly affected by the UAV ($y^{(m,n)}=1$) provide more significant information for localization, whereas unaffected pairs ($y^{(m,n)}=0$) may contribute limited information and even degrade the final localization accuracy. Therefore, a pair selection step is required before localization to identify the strongly affected BS-CPE pairs, so that the subsequent localization process can focus on the most informative observations.

During the inference of C-UDetNet, the CPU can additionally obtain the attention score of each BS-CPE pair. Since these scores reflect the relative contribution of different pairs to UAV detection, they can also serve as an effective indicator of pair selection for localization. Accordingly, a pair selection function $\mathcal{F}(\cdot)$ is introduced to select the BS-CPE pairs used for localization based on the attention scores
\begin{equation}
    \mathcal{S}_{\mathrm{loc}}=\mathcal{F}\!\left(\{a_{m,n}\}_{m=1,n=1}^{M,N}\right),
\end{equation}
where $\mathcal{S}_{\mathrm{loc}}$ denotes the set of selected BS-CPE pairs with $L$ elements, satisfying $0<L\leq M\times N$. After the effective pairs are determined, the CPU informs the corresponding pair indices to the associated BSs to initiate the subsequent individual localization process.

A straightforward solution is to select the top-$k$ pairs with the highest attention scores. However, according to \eqref{eq:att_score}, the attention scores $a_{m,n}$ are normalized by a softmax function, which ensures that the sum of all $M \times N$ scores equals 1. For different samples, the number of truly informative pairs ($y^{(m,n)}=1$) vary, and a fixed top-$k$ strategy may still retain some irrelevant pairs. To address this issue, an threshold $\sigma_{\mathrm{att}}$ is introduced to further remove pairs with low attention scores. Accordingly, the selection rule can be written as
\begin{equation}
\begin{aligned}
\mathcal{S}_{\mathrm{loc}}&= \mathcal{F}\!\left(\{a_{m,n}\}_{m=1,n=1}^{M,N}; k,\sigma_{\mathrm{att}}\right) \\
&= \left\{ (m,n)\,\middle|\, (m,n)\in \mathcal{S}_{\mathrm{top}\text{-}k},\ a_{m,n}\geq \sigma_{\mathrm{att}} \right\}.
\end{aligned}
\end{equation}
where $\mathcal{S}_{\mathrm{top}\text{-}k}$ denotes the set of pairs selected by the top-$k$ rule. Only the pairs that are ranked among the top-$k$ in attention scores and simultaneously satisfy the threshold requirement are retained for localization.

\subsection{Stage 1: Individual Localization}
\begin{figure*}[!t]
    \centering
    \includegraphics[width=0.95\linewidth]{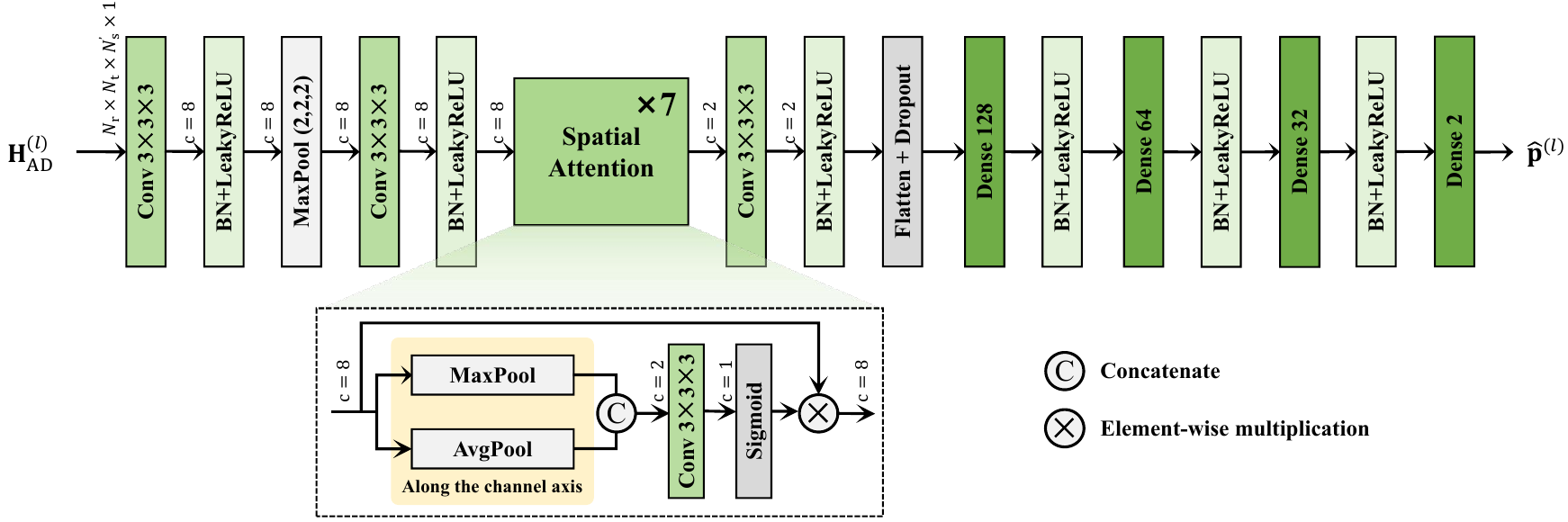}
    \caption{The NN architecture of the I-ULocNet for individual localization.}
    \label{fig:I-ULocNet}
    \vspace{-0.3cm}
\end{figure*}

Following the attention-guided pair selection operation, BSs are informed by CPU to execute individual localization on selected pairs. For each pair in $\mathcal{S}_\mathrm{loc}$, the UAV position $\hat{\mathbf{p}}^{(l)}$ is estimated from the corresponding CSI $\mathbf{H}^{(l)}$. Concurrently, compact CSI features $\mathbf{v}^{(l)}$ that preserve key propagation characteristics are extracted to enable subsequent cooperative sensing with minimal transmission overhead, thereby supporting further positioning enhancements. This stage includes a CSI preprocessing module and a localization NN, termed \underline{I}ndividual \underline{U}AV \underline{Loc}alization \underline{Net} (I-ULocNet). Formally, the individual localization is expressed as
\begin{equation}
    \hat{\mathbf{p}}^{(l)}, \mathbf{v}^{(l)} = g_\mathrm{I}(\mathcal{P}(\mathbf{H}^{(l)});\mathbf{\Phi}_\mathrm{I}),
\end{equation}
where $g_\mathrm{I}(\cdot)$ and $\mathbf{\Phi}_\mathrm{I}$ denote the I-ULocNet and its corresponding parameters, respectively. The CSI preprocessing module $\mathcal{P}(\cdot)$ converts the raw CSI into the angle-delay domain, producing $\mathbf{H}_\mathrm{AD}^{(l)}\in\mathbb{R}^{N_\mathrm{r}\times N_\mathrm{t}\times N_{\mathrm{c}}'\times 1}$, as described in Sec.~\ref{subsubsec:CSIpre}. I-ULocNet generates coarse position estimates and CSI features of each pair. A common I-ULocNet architecture with shared parameters is adopted for all BS-CPE pairs. The following describes the design of I-ULocNet.

\subsubsection{Input and Output of I-ULocNet} 
After preprocessing, I-ULocNet processes $\mathbf{H}_\mathrm{AD}^{(l)}$ to produce a coarse position $\hat{\mathbf{p}}^{(l)}$ and a lightweight CSI feature $\mathbf{v}^{(l)}$, which are the final and intermediate outputs of I-ULocNet, respectively. The ground-truth is the UAV's actual position $\mathbf{p}$. Specifically, the localization task demands the accurate inference of UAV-affected propagation paths from complex channels, followed by the effective mapping of these features to spatial coordinates within LAE airspace. Relying solely on coarse position estimates yields suboptimal performance. To address this, an intermediate NN layer of I-ULocNet (e.g., Dense or activation) produces a compact CSI feature $\mathbf{v}^{(l)} \in \mathbb{R}^{N_\mathrm{v}}$. The dimensionality $N_\mathrm{v}$ depends on the NN architecture, allowing flexible adjustment according to transmission resources. Collectively, the $L$ position estimates, associated CSI features, and pair indexes are transmitted for the subsequent cooperative localization stage.

\subsubsection{NN Architecture}
The NN architecture of I-ULocNet, illustrated in Fig.~\ref{fig:I-ULocNet}, comprises Conv3D, BN, MaxPool, Flatten, and Dense layers, along with several Spatial Attention (SA) blocks. The SA blocks, introduced in \cite{woo2018cbam}, aim to focus on informative regions of the input feature by generating a spatial attention map, computed via MaxPool, AveragePool (AvgPool), Conv3D, and Sigmoid operations. In individual localization, these blocks help extract UAV-relevant channel features more effectively, thereby improving accuracy.

To generate the spatial attention map, two pooling operations are first applied along the channel axis of the input feature $\mathbf{F}_\mathrm{in}$, and their outputs are concatenated to form a two-channel feature descriptor $\mathbf{F}_\mathrm{s}$
\begin{equation}
    \mathbf{F}_\mathrm{s}=[{\rm MaxPool} (\mathbf{F}_\mathrm{in}), {\rm AvgPool} (\mathbf{F}_\mathrm{in})].
\end{equation}
This descriptor emphasizes informative regions in the input. Then, $\mathbf{F}_\mathrm{s}$ is processed through a Conv3D layer followed by a Sigmoid activation to generate a spatial attention map $\mathbf{M}_\mathrm{s}$
\begin{equation}
    \mathbf{M}_\mathrm{s} = \rm Sigm({{\rm Conv3D} (\mathbf{F}_\mathrm{s})}).
\end{equation}
Finally, the output of the SA module is obtained through element-wise multiplication of the spatial attention map with the input feature map, expressed as
\begin{equation}
    \mathbf{F}_\mathrm{out} = \mathbf{F}_\mathrm{in} \odot \mathbf{M}_\mathrm{s}.
\end{equation}

To achieve a balance between transmission overhead and sensing performance, the compact CSI feature $\mathbf{v}^{(l)}$ is extracted from layers with low-dimensional outputs. Specifically, Dense and activation layers with output dimensions of 128, 64, and 32 are considered, corresponding to $N_\mathrm{v}\in\{128,64,32\}$. The I-ULocNet is trained using the mean square error (MSE) loss function to minimize the discrepancy between the estimated position and the ground-truth.

\subsection{Stage 2: Cooperative Localization}
\begin{figure*}[!t]
    \centering
    \includegraphics[width=0.95\linewidth]{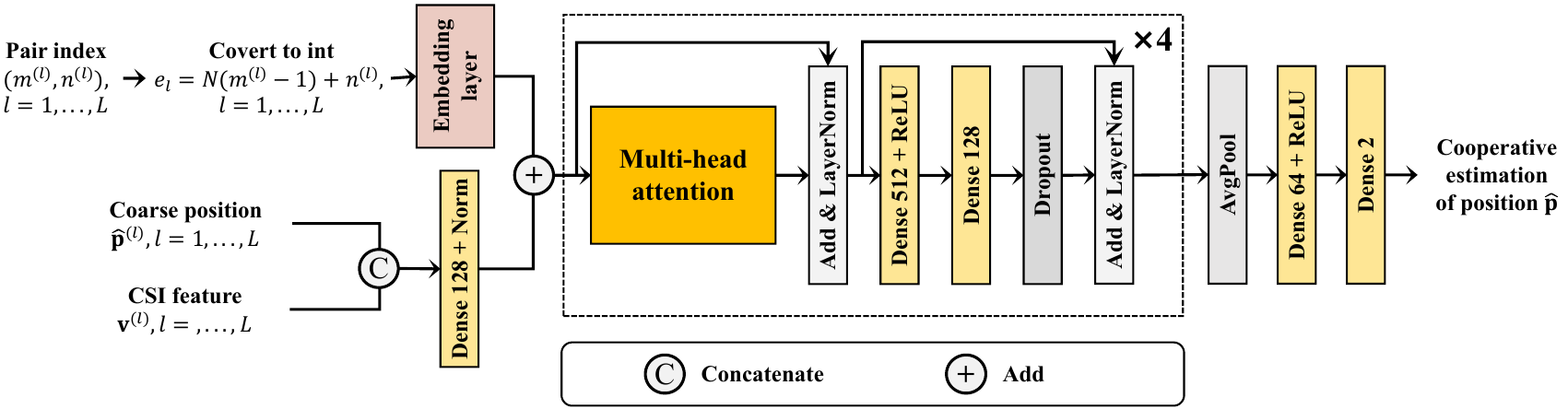}
    \caption{The NN architecture of the C-ULocNet for cooperative localization.}
    \label{fig:C-ULocNet}
\end{figure*}

Once individual localization is completed at $M$ BSs, the $L$ positioning estimates, CSI features, and corresponding pair indexes are transmitted to the CPU. In the cooperative localization stage, the CPU performs medium fusion, integrating received sensing data to produce a more precise UAV positioning estimate. Nevertheless, two challenges arise in this process. First, the number of UAV-affected BS-CPE pairs $L$ varies and depends on the UAV's location. Second, heterogeneity in the received sensing data further complicates the data fusion. To this end, a Transformer \cite{vaswani2017attention} is adopted as the backbone of the \underline{C}ooperative \underline{U}AV \underline{Loc}alization \underline{Net}, written as C-ULocNet. 
Let $\hat{\mathbf{P}} = {\left[\hat{\mathbf{p}}^{(1)}, \dots, \hat{\mathbf{p}}^{(L)}\right]}$, $\mathbf{V} = {\left[\mathbf{v}^{(1)}, \dots, \mathbf{v}^{(L)}\right]}$, and $\mathbf{e} = \left[e_1, \dots, e_L\right]$ denote the $L$ estimates, CSI features, and pairing indexes, respectively, where the $l$-th index is given by $e_l = N (m^{(l)}-1) + n^{(l)}$. Therefore, the cooperative localization is written as
\begin{equation}
    \hat{\mathbf{p}} = g_\mathrm{C}(\hat{\mathbf{P}},\mathbf{V},\mathbf{e};\mathbf{\Phi}_\mathrm{C}),
\end{equation}
where $g_\mathrm{C}(\cdot)$ and $\mathbf{\Phi}_\mathrm{C}$ denote the C-ULocNet and its corresponding parameters, respectively.

\subsubsection{Input and Output of C-ULocNet}
C-ULocNet leverages three sensing observations as input: positioning estimates, CSI features, and pair indexes. Each of them contributes complementary information for cooperative localization.
\begin{itemize}[leftmargin=*]
    \item \textbf{The coarse positioning estimates} provide spatial priors that constrain the search space and enhance fusion stability. 
    \item \textbf{The CSI feature vectors} capture essential propagation characteristics, such as angle-delay distributions, which are critical for fine-grained cooperative localization. 
    \item \textbf{The pair indexes} indicate the belonging BS-CPE pairs, providing geometry-aware context that allows C-ULocNet to model pair-specific propagation effects and perform adaptive compensation across multiple sensing nodes.
\end{itemize}
Through a dedicated learning process, C-ULocNet estimates the position of the UAV with enhanced accuracy.

\subsubsection{NN Architecture}
The architecture of C-ULocNet is shown in Fig.~\ref{fig:C-ULocNet}. The NN adopts a Transformer encoder-only framework, consisting of an input projection with an embedding layer, an encoder, and a localization head. The encoder includes four layers, each containing a multi-head self-attention (MHA) block, a feed-forward network (FFN), residual connections \cite{he2016deep}, and layer normalizations (LayerNorm). The overall learning process is illustrated as follows.

\begin{enumerate}[leftmargin=*,label=\textbf{\alph*.}]
    \item \textbf{Input projection and embedding layer.} First, the CSI feature matrix $\mathbf{V}$ is concatenated with the coarse positions matrix $\hat{\mathbf{P}}$. Then, the concatenated matrix is projected and normalized to produce sensing data embeddings $\mathbf{S}$, which is given by
    \begin{equation}
      \mathbf{S} = \mathrm{Norm}\left(\mathrm{Dense}([\hat{\mathbf{P}},\mathbf{V}])\right).
    \end{equation}
    Meanwhile, the pair indexes are converted to int and encoded as embeddings. Therefore, the input embeddings for the first encoder layer $\mathbf{X}$ are obtained by summing the sensing data embeddings and the positional embeddings
    \begin{equation}
      \mathbf{X}=\mathbf{S}+\mathrm{Embed}\left(\mathbf{e}\right).
    \end{equation}
    
    \item \textbf{Encoder.} The encoder module is composed of four identical layers, each following the standard Transformer structure and comprising two main components.
    \begin{itemize}[leftmargin=*]
        \item \textbf{A MHA block} whose output is combined with the input via a residual connection and LayerNorm. In this work, each MHA block has eight heads.
        \item \textbf{A two-layer FFN} followed by a Dropout layer. The output of FFN is added to its input through a residual connection and LayerNorm.
    \end{itemize}
    Consequently, the encoder layer can be formulated as
    \begin{gather}
        \mathbf{X}_\mathrm{a} = \mathrm{LayerNorm}(\mathbf{X} + \mathrm{MHA}(\mathbf{X})),\\
        \mathbf{X}_\mathrm{e} = \mathrm{LayerNorm}(\mathbf{X}_\mathrm{a} + \mathrm{Dropout}(\mathrm{FFN}(\mathbf{X}_\mathrm{a}))).
    \end{gather}
    
    \item \textbf{Localization head.} After four encoder layers, the outputs are aggregated by AvgPool over the token dimension to obtain a fixed-size global representation. The pooled vector is passed through several Dense layers to produce the final positioning estimate $\hat{\mathbf{p}}$
    \begin{equation}
        \hat{\mathbf{p}} = \mathrm{Dense}(\mathrm{ReLU}(\mathrm{Dense}(\mathrm{AvgPool}(\mathbf{X}_\mathrm{e})))).
    \end{equation}
\end{enumerate}

In summary, C-ULocNet leverages a Transformer encoder to integrate heterogeneous sensing data into a unified embedding space. Its self-attention mechanism aggregates complementary information from distributed BS-CPE pairs and captures long-range dependencies, thereby enhancing sensing robustness and accommodating variable $L$. Furthermore, the FFN, combined with residual connections and LayerNorm, stabilizes training and enables accurate, scalable cooperative localization.
During the training process, C-ULocNet is optimized via the MSE loss, and its localization performance is evaluated by the mean and cumulative distribution function (CDF) of the APE \cite{3GPP22137}. APE is defined as the Euclidean distance between $\hat{\mathbf{p}}$ and $\mathbf{p}$.

\begin{figure*}[!t]
    \centering
    \subfigure[The simulation layout of the LAE scenarios.]{
        \includegraphics[width=0.4\linewidth]{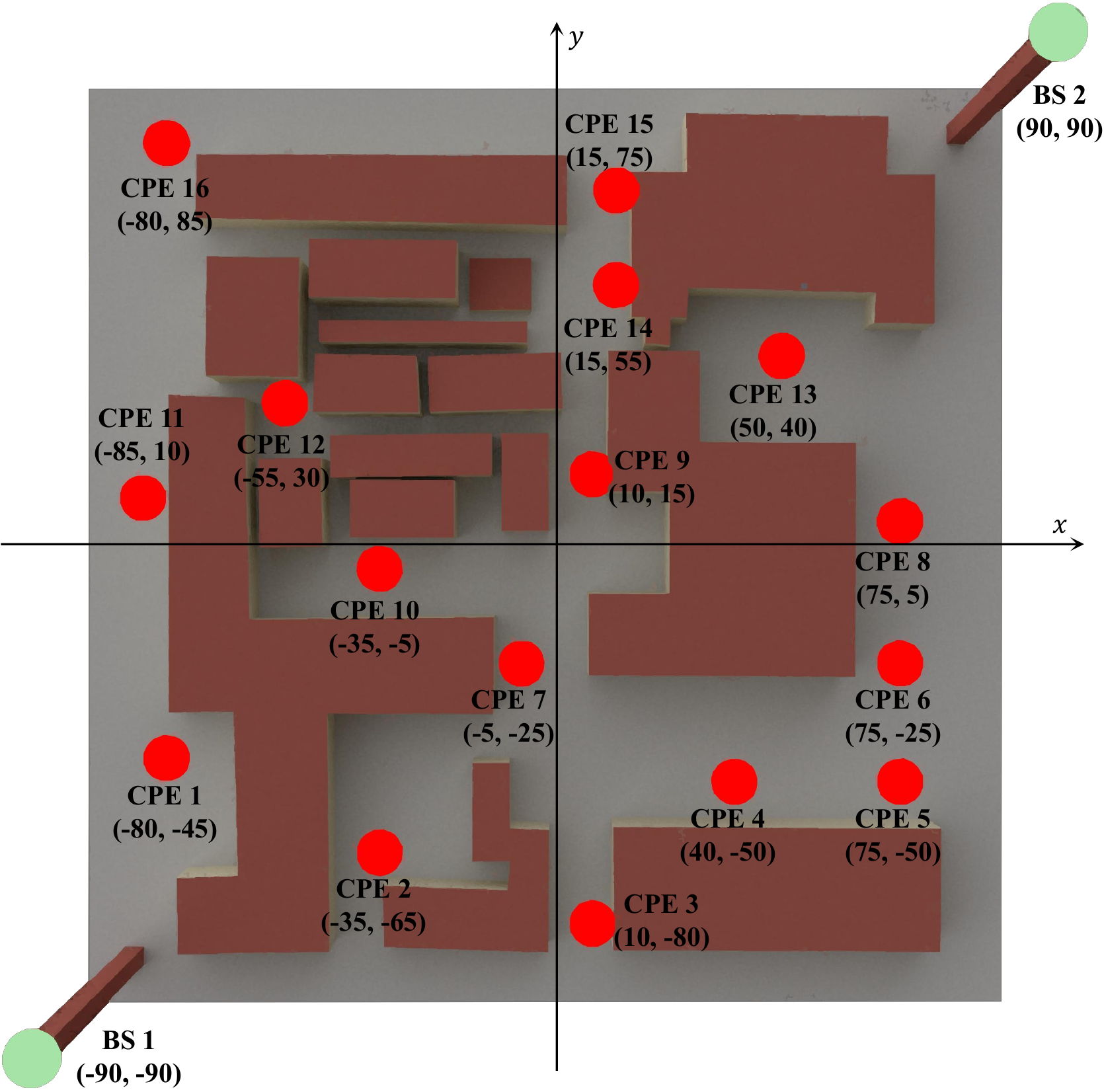}
        \label{fig:2Dwopaths}}
    \subfigure[Simulation of ray-tracing propagation paths. The black, blue, and green lines represent LoS, specular, and diffuse reflections, respectively.]{
        \includegraphics[width=0.47\linewidth]{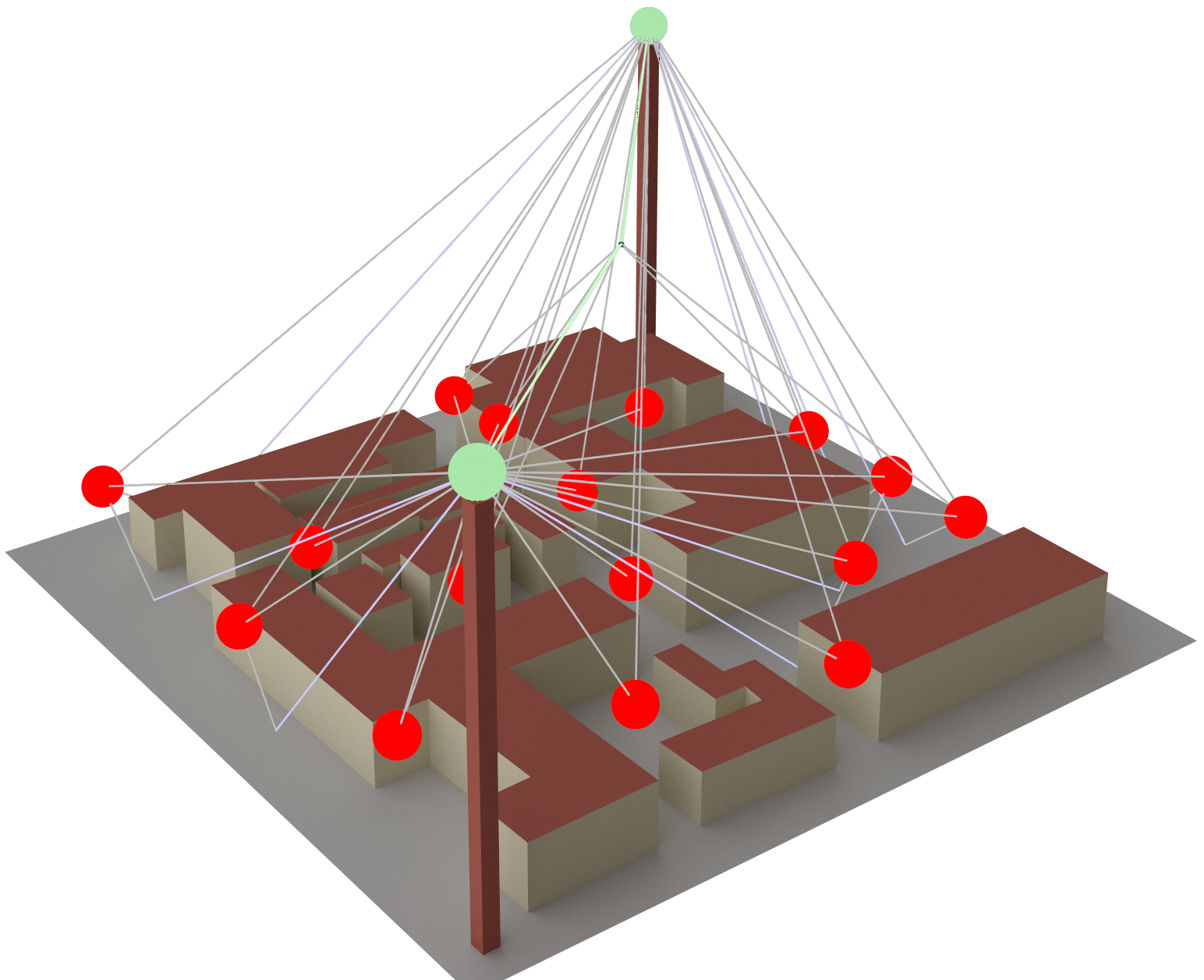}
        \label{fig:3Dwpaths}}
    \caption{The LAE scenario modeled with OSM, Blender, and Sionna.}
    \label{fig:Scenario_sionna}
    \vspace{-0.3cm}
\end{figure*}

\begin{figure*}[!t]
    \centering
    \subfigure[MDP]{
        \includegraphics[width=0.425\linewidth]{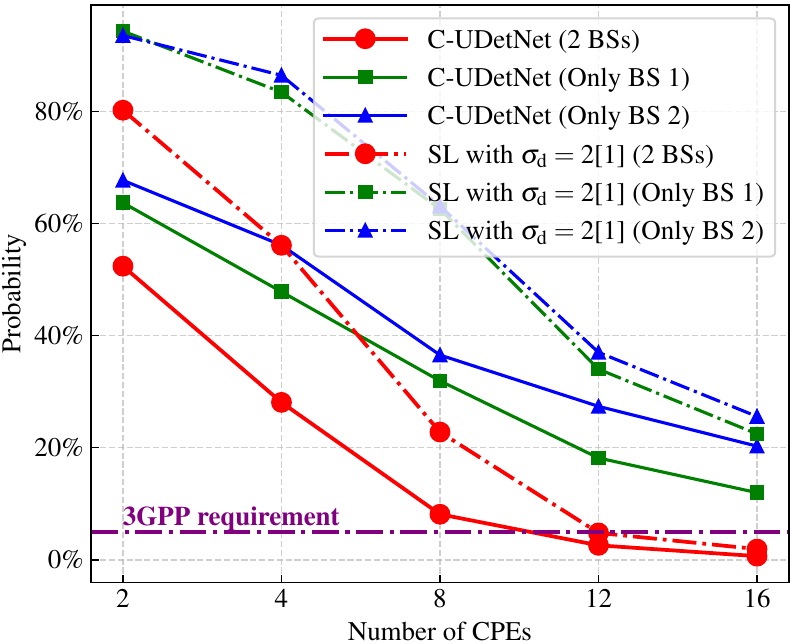}
        \label{fig:MDP}}
    \subfigure[FAP]{
        \includegraphics[width=0.43\linewidth]{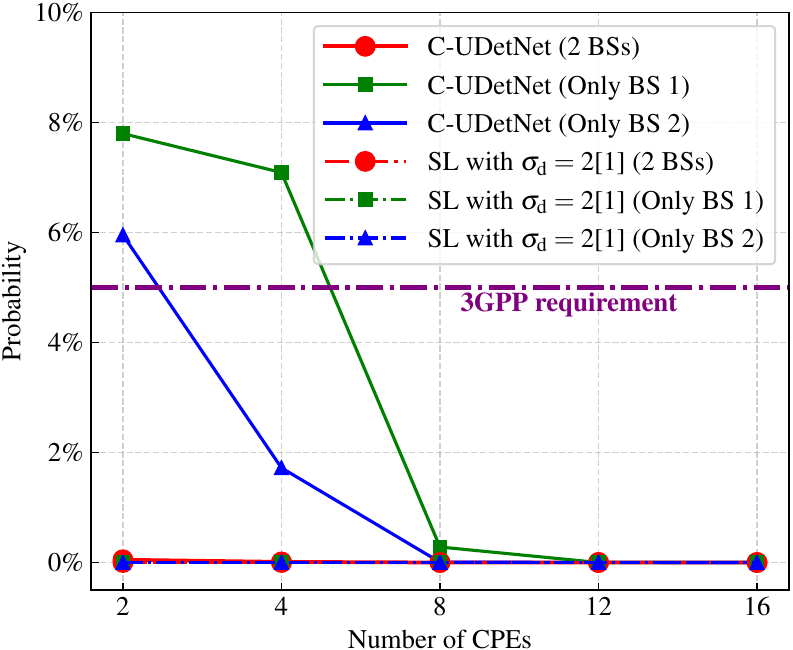}
        \label{fig:FAP}}
    \caption{Cooperative detection performance with different numbers of BSs/CPEs.}
    \label{fig:codet}
\end{figure*}

\begin{table}[!t]
\centering
\renewcommand{\arraystretch}{1.3}
\caption{Dataset segmentation of samples with/without UAV.}
\label{tab:dataset}
\begin{tabular}{p{1.8cm} p{1.8cm} p{1.8cm} p{1.8cm}}
\toprule
& \textbf{With UAV} & \textbf{Without UAV} & \textbf{Total samples} \\ 
\midrule
\textbf{Training} & 8,100 & 8,100 & \textbf{162,000} \\
\textbf{Validation} & 1,800 & 1,800 & \textbf{3,600} \\
\textbf{Testing} & 2,700 & 2,700 & \textbf{5,400}\\
\bottomrule
\end{tabular}
\end{table}

\section{Simulation Result}
\subsection{Simulation Settings}
\subsubsection{Scenario Configuration}
\label{subsubsec:sceneconfig}
The dataset is generated through ray-tracing, using OpenStreetMap (OSM), Blender, and Sionna RT \cite{hoydis2023sionna}. A 200~m $\times$ 200~m area of Southeast University Sipailou Campus is first extracted from OSM and imported into Blender, where building materials such as concrete, marble, and brick are assigned. The scenario includes $M=2$ BSs and $N=16$ CPEs positioned at heights of $z_\mathrm{bs}^1 = z_\mathrm{bs}^2 = 100$~m and $z_\mathrm{cpe}^n=18$~m, $n = 1, \dots,16$, respectively. Their x-y plane coordinates are illustrated in Fig.~\ref{fig:2Dwopaths}. The UAV is modeled as a metallic cube located at $\mathbf{p}=(x,y,60)$, where $x,y \sim \mathcal{U}[-75,75]$.

For the MIMO-OFDM system, each BS is equipped with $N_\mathrm{r}=64$ antennas and each CPE with $N_\mathrm{t}=4$ antennas, respectively. The choice of $N_\mathrm{t}=4$ is aligned with commercial FWA CPE configurations, since 5G FWA products with 4x4 MIMO are already commercially available. In addition, FWA planning studies indicate that one site may serve hundreds of households, suggesting the practical potential of the proposed method in real-world FWA deployment scenarios. The carrier frequency is 2.8 GHz with a 20 MHz bandwidth, 30 kHz subcarrier spacing, and $N_\mathrm{c}=512$ subcarriers. During ray-tracing, the wireless propagation channel accounts for LoS and first-order reflections comprising both specular and diffuse components, as shown in Fig.~\ref{fig:3Dwpaths}. In the simulation, due to the scene geometry and the strict conditions required for specular reflection, most UAV-affected paths are dominated by diffuse reflections, which generally have lower power. Accordingly, the CSI logarithmic compression step in \eqref{eq:prepro_log} is helpful to reveal these low-energy paths.

\subsubsection{CSI Generation}
The dataset is divided into samples with and without UAV, and further split into training, validation, and testing sets, as summarized in Table~\ref{tab:dataset}. For the UAV detection task, $y^\mathrm{s}$ is set to 1 whenever a UAV exists in the scenario, while the pair labels $y^{(m,n)}$ is derived from the path type provided by Sionna RT. As described in Sec.~\ref{subsubsec:sceneconfig}, UAV-affected paths are typically caused by diffuse reflections. Therefore, $y^{(m,n)}$ is set to 1 only when the channel between a BS-CPE pair contains such reflections. These pair labels are not involved in the training of I-UDetNet or C-UDetNet, but are used only to evaluate the effectiveness of attention score learning in detection and the pair selection strategy for localization. For the UAV localization task, the ground-truth is the actual UAV position denoted by $\mathbf{p}$.

\subsubsection{Data Augmentation}
The channel parameters in \eqref{eq:ori_channel} are obtained from path instances computed by the PathSolver function in Sionna RT. CSI samples from a fully static scenario are nearly constant, which may cause overfitting during training. Therefore, a random phase shift is introduced to each path based on \eqref{eq:ori_channel} to increase data diversity
\begin{equation}
     H_{i,j}(f) = \sum_{p=1}^{N_{\mathrm{p}}} \alpha_{i,j,p} e^{-j2\pi f \tau_{p}}e^{j\theta_p},
\end{equation}
where $\theta_p \sim \mathcal{U}[0,2\pi)$ expresses the random phase shift.

\subsubsection{Training Configuration}

For CSI preprocessing, only the first $N_{\mathrm{c}}'=64$ elements along the delay dimension are retained. All NNs are optimized using Adam optimizer. Specifically, the jointly trained I-UDetNet and C-UDetNet employs an initial learning rate of $1 \times 10^{-4}$, a batch size of 64, and is trained for 200 epochs. I-ULocNet and C-ULocNet use an initial learning rate of $1 \times 10^{-3}$ and a batch size of 32, with training durations of 1,000 and 300 epochs, respectively. To enhance training stability and mitigate overfitting, early stopping and learning rate decay strategies are applied across all NNs.

\subsection{Performance Evaluation of Cooperative Detection}

\subsubsection{Detection Performance}

\begin{figure}[!t]
    \centering
    \subfigure[Sample A.]{
        \includegraphics[width=1\linewidth]{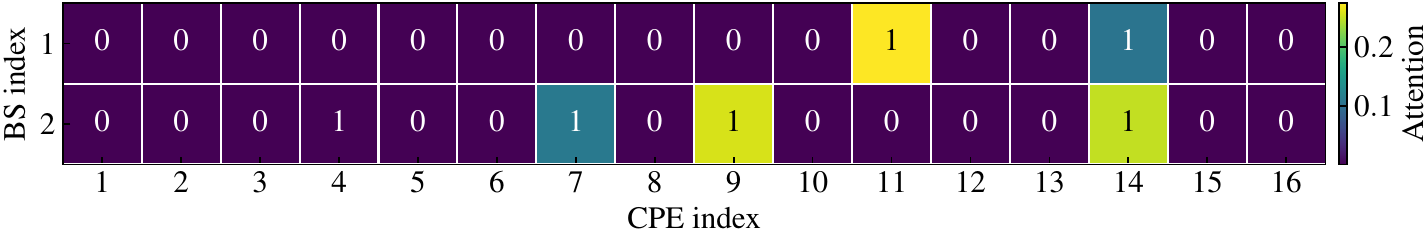}
        \label{fig:att22}}
    \subfigure[Sample B.]{
        \includegraphics[width=1\linewidth]{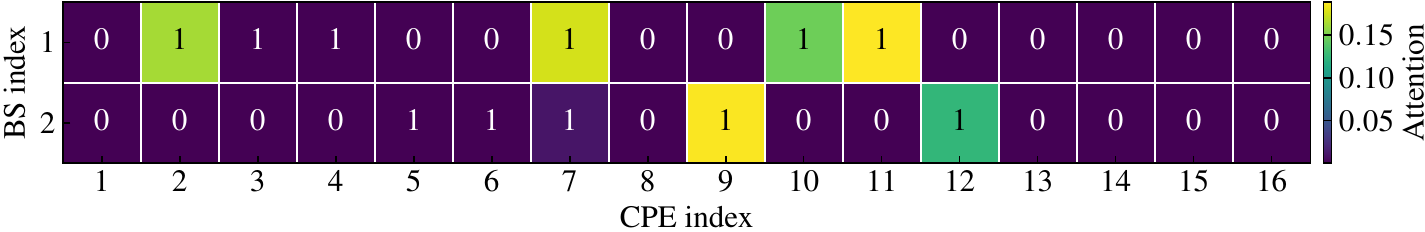}
        \label{fig:att25}}
    \caption{Visualization of attention scores and true pair labels. In each grid, the color intensity indicates the attention weight, while the 0/1 entry denotes the pair label $y^{(m,n)}$. Pairs with higher attention weights contribute more to the NN decision, and the consistency between the learned attention distribution and the true pair labels demonstrates the effectiveness of the proposed framework.}
    \vspace{-0.15cm}
    \label{fig:attention scores}
\end{figure}

\begin{figure}[!t]
    \centering
    \includegraphics[width=0.95\linewidth]{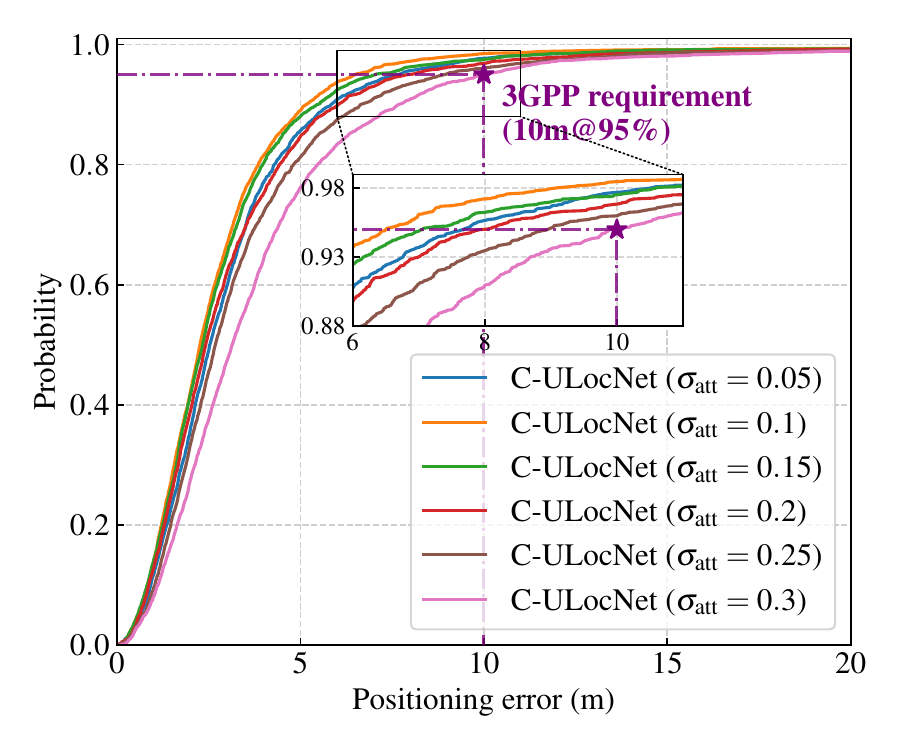}
    \caption{With $k=10$ and varying $\sigma_\mathrm{att}$, the CDF of APE performance in the proposed two-stage localization framework.}
    \label{fig:CDF_NN_x20}
    \vspace{-0.15cm}
\end{figure}

\begin{table}[]
\centering
\renewcommand{\arraystretch}{1.3}
\setlength{\tabcolsep}{9pt}
\caption{The reliability of attention-guided pair selection strategy with vary $\sigma_\mathrm{att}$.}
\label{tab:att_guided_sel}
\begin{tabular}{ccccc}
\toprule
\textbf{$\sigma_\mathrm{att}$} & \textbf{TP} & \textbf{FP} & \textbf{FN} & \textbf{Range of $L$} \\
\midrule
0.05 & 26,334 & 347 & 21,703 & {[}1,10{]} \\
0.1 & 23,861 & 116 & 24,176 & {[}1,8{]} \\
0.15 & 21,189 & 60 & 26,848 & {[}1,6{]} \\
0.2 & 17,634 & 57 & 30,403 & {[}1,4{]} \\
0.25 & 14,278 & 56 & 33,759 & {[}1,3{]} \\
0.3 & 11,752 & 56 & 36,285 & {[}1,3{]} \\
True pair label & \textbackslash{} & \textbackslash{} & \textbackslash{} & {[}1,15{]} \\
\bottomrule
\end{tabular}
\smallskip
\begin{flushleft}
    \footnotesize
    * TP=True positive, FP=False positive, FN=False negative.
\end{flushleft}
\vspace{-0.3cm}
\end{table}

In the cooperative detection stage, the MDP and FAP are used to assess the accuracy of UAV detection. 
As a benchmark for validating the effectiveness of the proposed framework, the performance requirements specified in 3GPP TR 22.870 \cite{3GPP22870} are considered, where both the MDP and FAP are required to be below 5\%. In addition, the supervised learning (SL)-based I-UDetNet with the thresholding-based decision rule proposed in \cite{zhang2026icc} is included as a comparison scheme, denoted by SL with $\sigma_\mathrm{d}$.

Fig.~\ref{fig:codet} shows the performance of the proposed cooperative detection framework under different BS/CPE configurations. The proposed method consistently achieves competitive detection performance, although its training process does not rely on explicit pair labels. This result indicates that the joint I-UDetNet and C-UDetNet scheme can adaptively identify the UAV-affected pairs, and effectively exploit the complementary information provided by multiple pairs to infer the presence of UAV. In particular, with $M=2$ BSs and $N=16$ CPEs, the proposed method achieves an MDP of 0.63\% and an FAP of 0.00\%. The 95\%-confidence interval of MDP is [0.37\%, 1.01\%], and that of FAP is [0.00\%, 0.13\%]. Both metrics satisfy 3GPP requirements, which supports the practical feasibility of the proposed method for cooperative UAV detection.

Fig.~\ref{fig:codet} also illustrates the impact of the BSs/CPEs numbers on the detection performance. As the number of participating pairs increases, the MDP decreases, indicating that multi-view observations provide complementary sensing information and help mitigate sensing blind spots. Besides, Fig.~\ref{fig:FAP} shows that reducing the number of participating pairs leads to a higher FAP. When more pairs are available, random non-UAV-related fluctuations can be effectively suppressed by the consistency across multiple views. In contrast, with only a few pairs, the detection result becomes much more sensitive to channel variations, therefore causing more overconfident false alarms.

Overall, these results suggest that cooperative sensing performance depends not only on the number of observations and sensing coverage, but also on the reliability of the participating views. Additionally, for low-altitude UAV intrusion detection, minimizing MDP is of higher priority than lowering FAP, since missed detections are unacceptable.

\begin{figure*}[!t]
    \centering
    \subfigure[Mean APE.]{
        \includegraphics[width=0.47\linewidth]{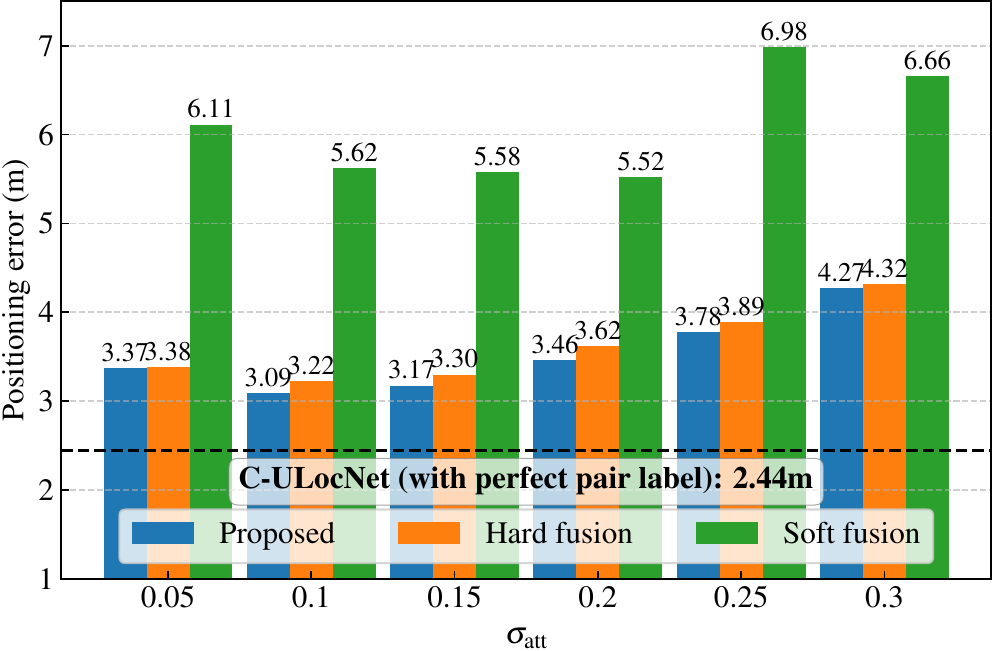}
        \label{fig:mean_nn_avg}}
    \subfigure[95\%-confidence APE.]{
        \includegraphics[width=0.485\linewidth]{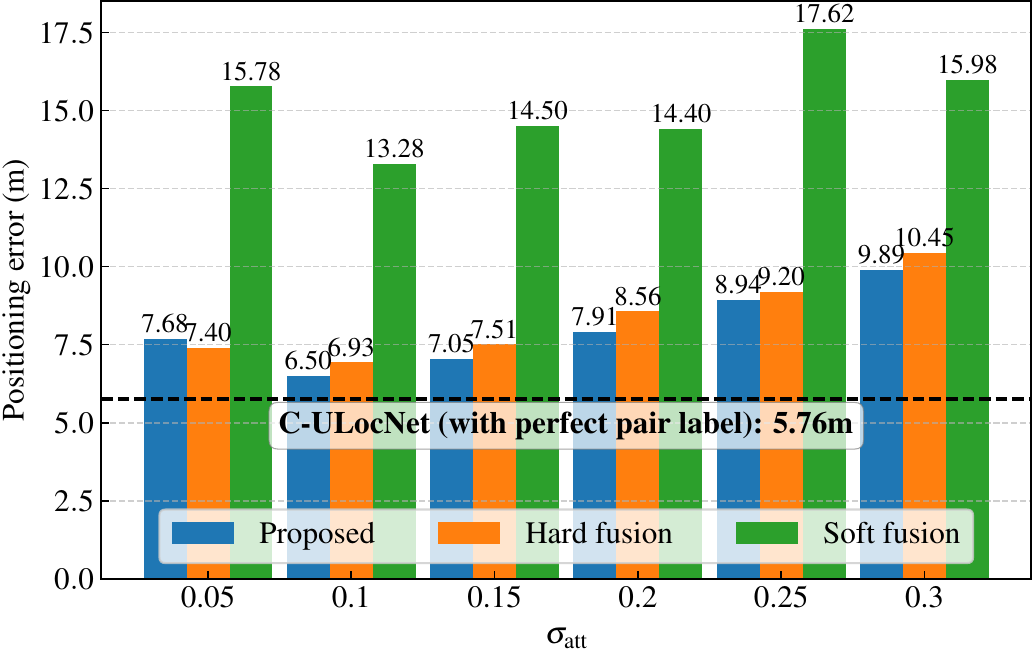}
        \label{fig:cdf95_nn_avg}}
    \caption{With varying $\sigma_\mathrm{att}$, APE comparison among proposed framework, hard fusion, and soft fusion.}
    \label{fig:nn_vs_avg}
\end{figure*}

\subsubsection{Attention Score Visualization}

To verify the effectiveness of learned attention scores, Fig.~\ref{fig:attention scores} visualizes the attention distribution over all pairs. Specifically, two samples are shown as examples, where the color of each grid indicates the attention score, and the number inside each grid denotes the pair label $y^{(m,n)}$. It can be observed that grids with relatively high attention scores are generally associated with positive pair labels, suggesting that the proposed attention mechanism can adaptively focus on the BS-CPE pairs affected by the UAV. Moreover, the Pearson correlation coefficient between the true pair labels and the attention scores is 0.66, further confirming the moderately strong consistency between them.

Nevertheless, not all positive BS-CPE pairs are assigned high attention scores. This phenomenon may be related to the varying degrees of CSI perturbation induced by the UAV across different pairs. To reduce unreliable decisions, the NN tends to assign higher attention weights to pairs with more pronounced UAV-induced channel variations. In addition, since the attention scores are normalized to sum to one, the attention allocated to each positive pair may decrease when the number of positive pairs becomes large.

\subsection{Performance Evaluation of Cooperative Localization}

The cooperative localization performance is evaluated against four schemes.
\begin{itemize}[leftmargin=*]
    \item \textbf{3GPP requirements.} A benchmark specified in 3GPP TS 22.870 \cite{3GPP22870} is used to validate the framework. Concretely, at the 95\% confidence level, the APE should be below 10~m.

    \item \textbf{C-ULocNet with perfect pair labels.} This scheme assumes that ground-truth pair labels ($y^{(m,n)}$) are utilized for pair selection, aiming to verify the performance of the proposed attention-guided selection method.
    
    \item \textbf{Hard fusion.} A baseline in which the $L$ positioning estimates obtained in individual localization stage are averaged as final result. This scheme serves to verify the effectiveness of the Transformer-based cooperative localization.

    \item \textbf{Soft fusion.} A baseline in which $L$ raw CSI are directly uploaded to CPU and fused using a Transformer. This scheme is used to demonstrate the effectiveness of the two-stage medium-fusion strategy.
\end{itemize}

\begin{figure}[!t]
    \centering
    \includegraphics[width=0.95\linewidth]{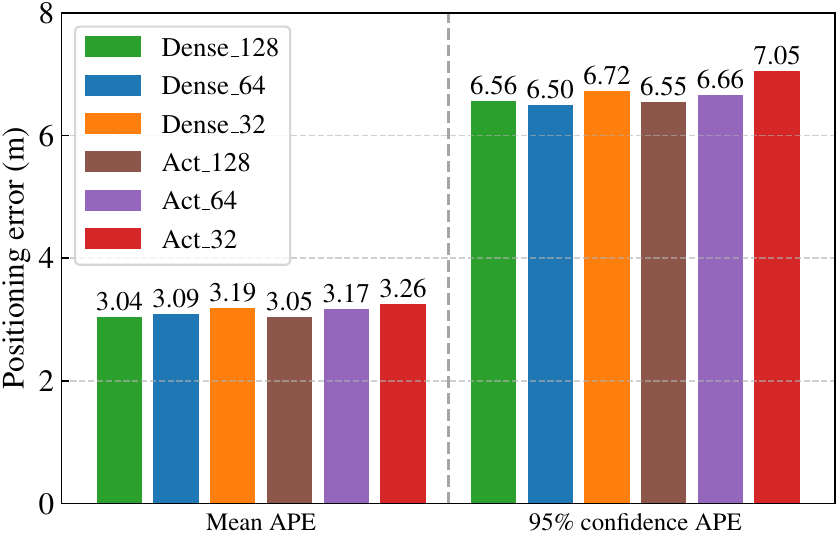}
    \caption{With varying CSI features, APE performance in cooperative localization under the configuration of $k=10$, $\sigma_\mathrm{att}=0.1$.}
    \label{fig:mean_cdf95_lys}
\end{figure}

\begin{figure*}[!t]
    \centering
    \subfigure[A single pair.]{
        \includegraphics[width=0.315\linewidth]{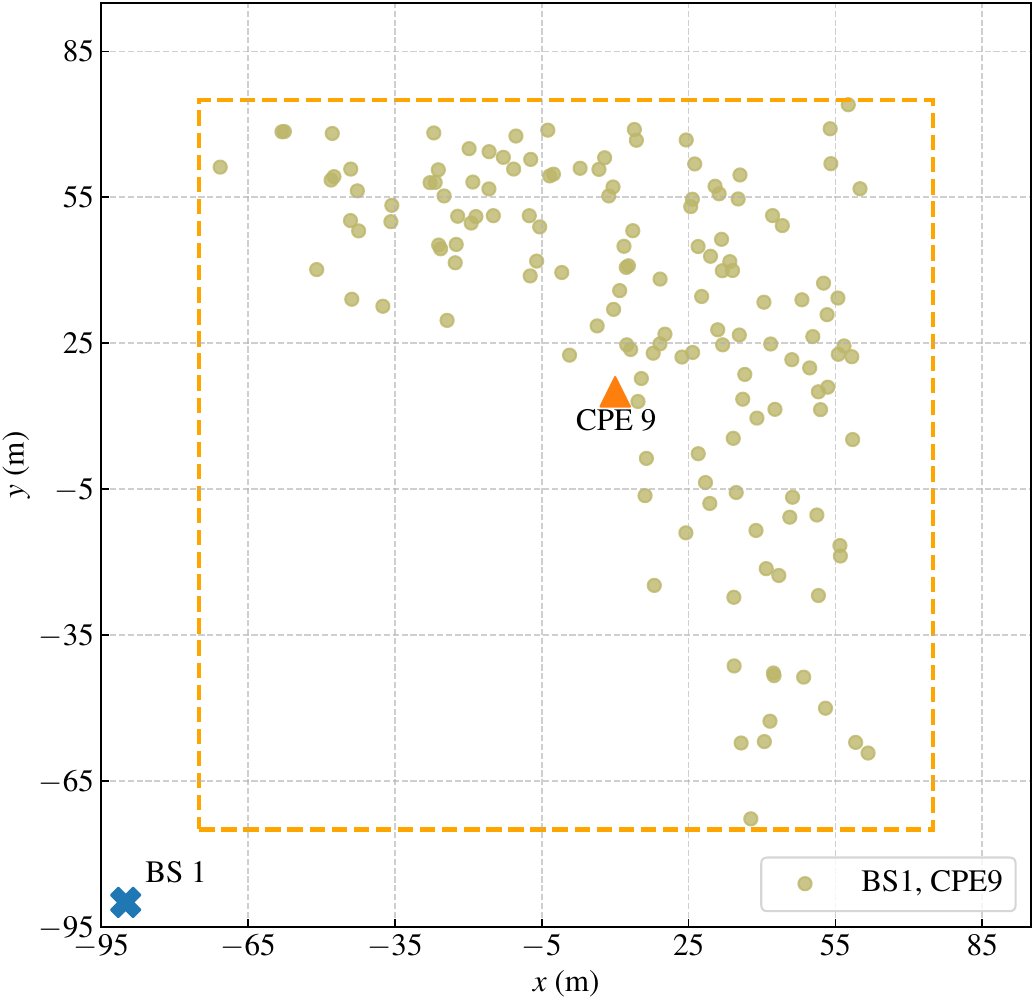}
        \label{fig:BS1CPE9dis}}
    \hfill
    \subfigure[Two BSs with a single CPE.]{
        \includegraphics[width=0.315\linewidth]{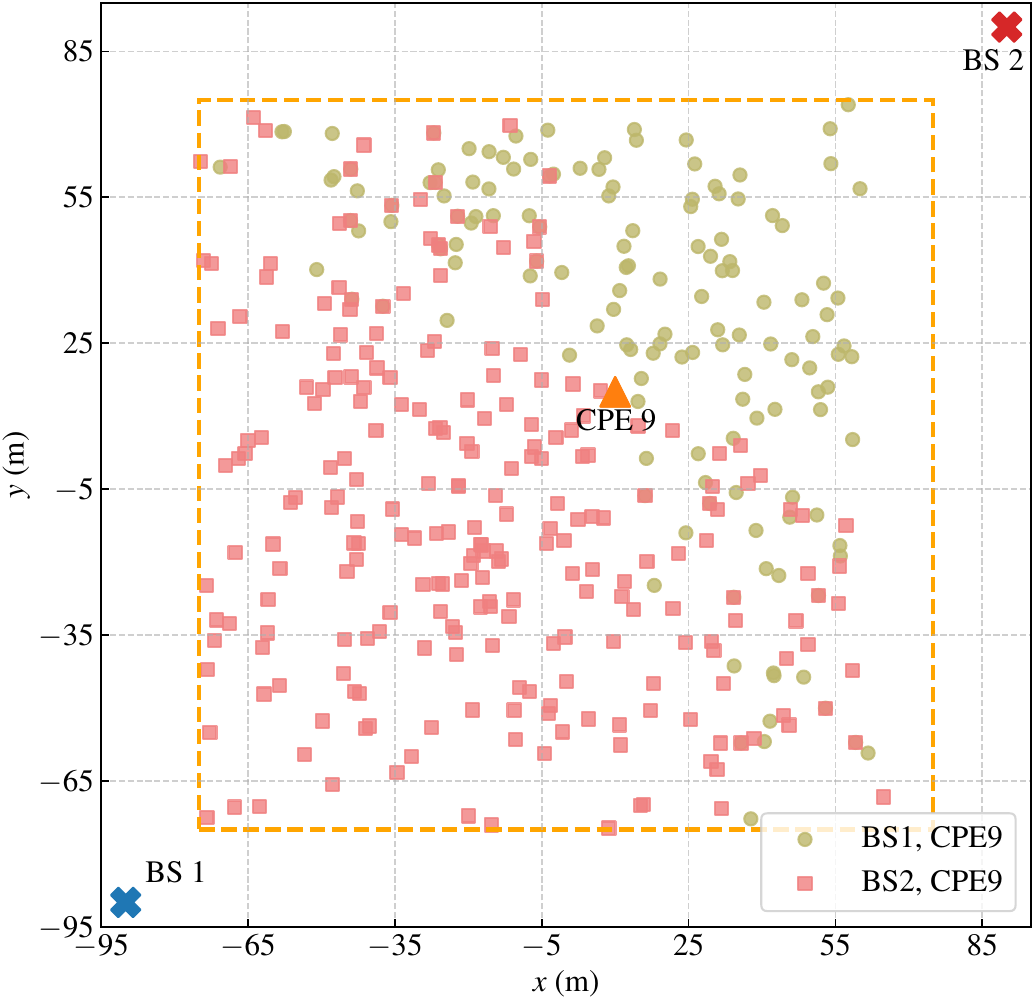}
        \label{fig:2BSdis}}
    \hfill
    \subfigure[A single BS with several CPEs.]{
        \includegraphics[width=0.315\linewidth]{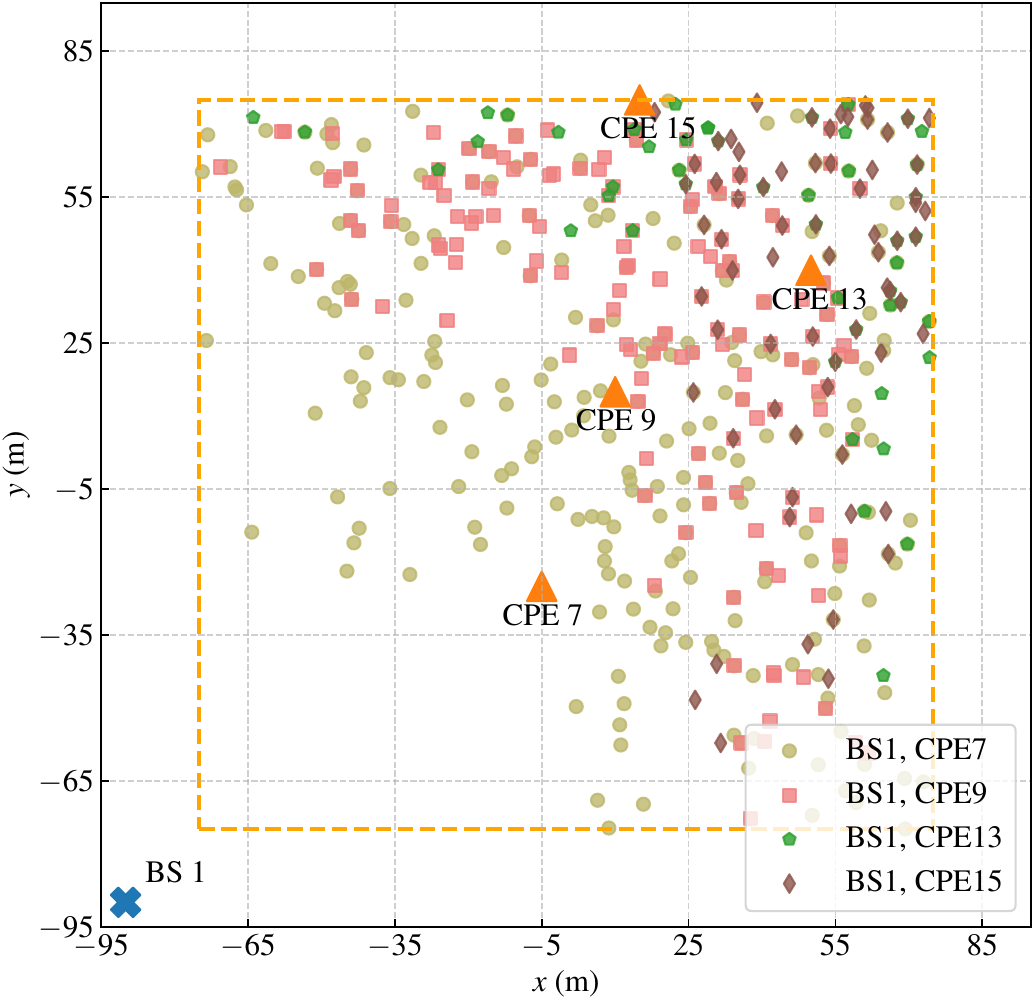}
        \label{fig:4CPEdis}}
    \caption{Sensing region visualizations of the samples from the testing dataset. Multiple BSs provide complementary coverage that reduces blind areas, while multiple CPEs produce overlapping regions that enhance robustness.}
    \label{fig:distribution}
\end{figure*}

\subsubsection{Reliability of Attention-Guided Pair Selection}
We first evaluate the attention-guided pair selection under different $\sigma_\mathrm{att}$ with $k=10$. Since the sum of attention scores equals to one, when $k$ is sufficiently large, the number of selected pairs $L$ is mainly determined by $\sigma_\mathrm{att}$. We set $\sigma_\mathrm{att} \in \{0.05, 0.1, 0.15, 0.2, 0.25, 0.3\}$, and Tab.~\ref{tab:att_guided_sel} compares the selected pair set $\mathcal{S}_{\mathrm{loc}}$ with the ground-truth pair labels over the entire training dataset in terms of TP, FP, and FN. As $\sigma_\mathrm{att}$ increases, fewer TP pairs are retained, while FP pairs are also reduced, revealing the tradeoff of the selection strategy. Retaining sufficient TP pairs is important for localization, whereas excessive FP pairs may introduce irrelevant or misleading information. Therefore, these results not only reflect the effectiveness of the attention-guided selection scheme, but also help explain the following localization performance under different threshold settings.

\subsubsection{Localization Performance}
First, the proposed framework is compared with three benchmark schemes. Unless otherwise specified, the CSI features in the individual localization stage are extracted from the Dense layer with $N_\mathrm{v}=64$. Fig.~\ref{fig:CDF_NN_x20} presents the CDF of the APE under different $\sigma_\mathrm{att}$. Under all considered threshold settings, C-ULocNet satisfies the 3GPP requirement of 10~m APE at the 95\% confidence level. Specifically, when $\sigma_\mathrm{att}=0.1$, the mean and 95\%-confidence APE are 3.09~m and 6.50~m, respectively. Moreover, the performance achieved by attention-guided pair selection remains close to that obtained with true pair labels, indicating that the attention scores learned in the detection stage are effective for select informative pairs. This also suggests the potential for a tight integration of UAV detection and localization tasks in practical deployment. Compared with hard and soft fusion schemes, the proposed two-stage medium fusion framework provides further performance gains, reducing the 95\%-confidence APE by 0.43~m and 6.78~m, respectively.

The influence of $\sigma_\mathrm{att}$ on APE is further illustrated in Fig.~\ref{fig:nn_vs_avg} and can be interpreted together with Tab.~\ref{tab:att_guided_sel}. As $\sigma_\mathrm{att}$ decreases, the mean and 95\%-confidence APE in Fig.~\ref{fig:nn_vs_avg} generally decrease, indicating improved localization performance. This trend can be attributed to the increased number of selected pairs $L$, which provides richer observations and enhances prediction accuracy. Consequently, NNs can exploit sufficient information for accurate UAV localization. Nevertheless, a slight performance degradation is observed when $\sigma_\mathrm{att}=0.05$. A possible reason is that the number of falsely selected pairs increases under this setting, introducing misleading information that degrades localization accuracy especially in cooperative stage.

Finally, we investigate the effect of CSI feature dimensionality on cooperative localization. In the simulation, CSI features with length $N_\mathrm{v}\in\{128,64,32\}$ are extracted from either a Dense layer or an activation layer in I-ULocNet. As shown in Fig.~\ref{fig:mean_cdf95_lys}, all feature settings achieve competitive cooperative localization performance, with the mean APE around 3.04 to 3.26~m and the 95\%-confidence APE around 6.50 to 7.05~m. These results suggest that the proposed I-ULocNet can effectively compress CSI into compact feature representations that preserve the key semantics required for localization while discarding redundant information. This indicates that compact CSI features are sufficient for cooperative localization, which helps reduce the transmission overhead between BSs and the CPU in practical deployment.

In summary, the choice of $\sigma_\mathrm{att}$ plays an important role in balancing localization accuracy and transmission overhead. A smaller threshold allows more pairs to participate in localization, thereby enhancing localization accuracy through richer CSI observations. However, it may also introduce unrelated pairs that hinder the localization process and consume additional transmission resources between BSs and the CPU. Therefore, the pair selection threshold $\sigma_\mathrm{att}$ should be carefully selected in practical deployment to achieve a desirable localization accuracy.

\subsection{Analysis and Discussions}
\label{subsec:Analysis and Discussions}
In this section, the sensing regions of BS-CPE pairs are visualized and used to analyze the cooperation benefits. 
\begin{itemize}[leftmargin=*]
    \item \textbf{Sensing region of each pair.} Fig.~\ref{fig:distribution} illustrates the spatial distribution of UAV samples for which $y^{(m,n)} = 1$ (i.e., CSI is perturbed by the UAV). For clarity, only some representative pairs are shown. As depicted in Fig.~\ref{fig:BS1CPE9dis}, empirical observations indicate that the UAV-detectable region of a pair tends to lie outside the rectangular area whose diagonal is defined by the segment between the BS and the CPE. Moreover, the sample distribution within the sensing region is non-uniform, with fewer samples observed in areas farther away from the pair.
    
    This behavior arises because the UAV operates at an altitude between the BS and CPE, establishing a dominant two-segment reflection path: from the CPE to the UAV, then from the UAV to the BS. For both segments to meet the geometric constraints, the UAV’s horizontal position must lie outside the BS-CPE rectangular region. Furthermore, long-distance propagation incurs severe path loss, resulting in sparse samples in such regions. These analysis of sensing geometries provides valuable insights for optimizing node deployment in practice.
    
    \item \textbf{Advantages of cooperative sensing.} The variability in sensing regions across BS-CPE pairs motivates us to adopt the cooperative frameworks to mitigate pair-specific blind zones. As illustrated in Fig.~\ref{fig:2BSdis} and Fig.~\ref{fig:4CPEdis}, multiple BSs provide complementary coverage that fills gaps left by individual pairs, whereas multiple CPEs often yield overlapping sensing regions when their geometries relative to the BSs are similar. Consequently, cooperative schemes not only reduce sensing blind spots but also enhance robustness by incorporating multiple sensing nodes.
\end{itemize}

\section{Conclusion}
In this paper, we propose AI-based, two-stage cooperative UAV sensing frameworks that leverage uplink CSI from multiple BS-CPE pairs for LAE supervision. For UAV detection tasks, CSI feature of each BS-CPE pair is first extracted and then cooperatively fused through an attention-based pooling mechanism to determine UAV presence. The learned attention scores further guide the selection of UAV-affected pairs with top-$k$ and thresholding strategy, providing significant information for UAV localization. In localization tasks, a medium-fusion strategy is adopted, in which individual position estimates and CSI features are first generated and subsequently integrated to produce a more accurate UAV position estimate. Establishing with existing BSs and CPEs, the proposed framework achieves substantial cooperative gains and reduces sensing data transmission overhead through its two-stage scheme. Simulation results show that cooperative frameworks significantly improve sensing coverage and reliability, achieving only 0.63\% missed detections and a positioning error of 6.50~m at the 95\% confidence level. These results suggest that the AI-based sensing framework, which incorporates FWA and BSs, provides a practical and accurate solution for LAE UAV monitoring.

\bibliographystyle{IEEEtran}
\bibliography{IEEEabrv,reference}

\end{document}